# Air Observe System*

**Alexander Bolonkin**

***C&R, 1310 Avenue R, #6-F, Brooklyn, NY 11229, USA. aBolonkin@junol.com, http://Bolonkin.narod.ru***

**Abstract**

This manuscript contains a description and basic principles for observing inaccessible areas using low cost, easily deployed equipment. The basic premise is to suspend a tiny video camera at an altitude of 10 -200 meters over the area to be surveyed. The TV camera supports at altitude by wind or ballom. The technical challenges regard the means by which the camera is suspended. Such a system may be used by military or police forces, or by civil authorities for rescue missions or assessment of natural disasters. The method may be further developed for military applications by integrating the surveillance task with deployment of munitions.

**Contents:**

# 1. Introduction.

### 1.1. Historical Perspective

From 1993-2000 the Defense Advanced Research Projects Agency (DARPA) spent 35 millions dollars for Micro Air Vehicle (MAV) research. Micro aircraft, no larger than a small bird, are already showing promise in reconnaissance roles by flying with video cameras and returning live pictures. At present time the Air Force and Army continue Micro Air Vehicle (MAVs) research and development for assisting ground soldiers with non-line-of-sight reconnaissance. Unfortunately, after 10 years of development and spending hundreds of millions of dollars, we still do not have a MAV suitable for reliable, sustainable close-in surveillance. The reason is that the MAV method is fundamentally limited for this role. It is impossible to use when the wind is strong. As fuel capacity is limited, observation times are very limited. An enemy can fairly easily see and avoid (or destroy) the MAV because the MAV must flights at low altitude when using small, lightweight cameras and optics. The soldier may require special training for control and operation of the MAV, especially if it does not have an autopilot.

We propose have a method based upon a simple device, which does not have these shortcomings and in some cases, may be more efficient than a MAV.

### 1.2. Short Description of the Micro Air Observer (MAO).

The basic approach for this method is to suspend a very small, essentially invisible, Micro Air Observer (MAO) or (SAO) -Suspended Air System at a controlled altitude over an area to be investigated. The MAO device includes: (1) aerial support device (kite, air balloon) located at a relatively high altitude (e.g. 1000 m) which is connected by a thin fiber cable and thin electric wire to anchors (and a battery) located at the Earth's surface; (2) a micro video camera (and microphone) at the low altitude (100-200m) connected to the support device by thin fiber cable and wire; (3) support electronics including a transceiver, radio control, and small battery; (4) control and observation ground station for the soldier (or operator). Optionally, the MAO may also contain a self-destructor. The entire device may be packaged in a canister and dispensed from an aircraft or artillery shell. Most of the devices required to build the MAO exist off-the-shelf and have a combined weight 20-90 grams and volume of 10-50 cubic centimeters.

There are several possible launch methods and conditions for the MAO:

a) The MAO is launched from aircraft. The aircraft dispenses the MAO canister into a given area (fig.1). The canister is opened at a given altitude, the kite (balloon) is opened, the anchor falls on the ground and stabilizes the kite (balloon). The kite deploys the camera to the desired altitude and begins operation. The kite may require additional anchors (fig.2) to ensure that the MAO will have the *same position for any wind direction*. The schematic design of the canister and anchor are shown in figs. 3 and 4 respectively. Fig.5 show the MAO being launched from an air launched cruise missile.

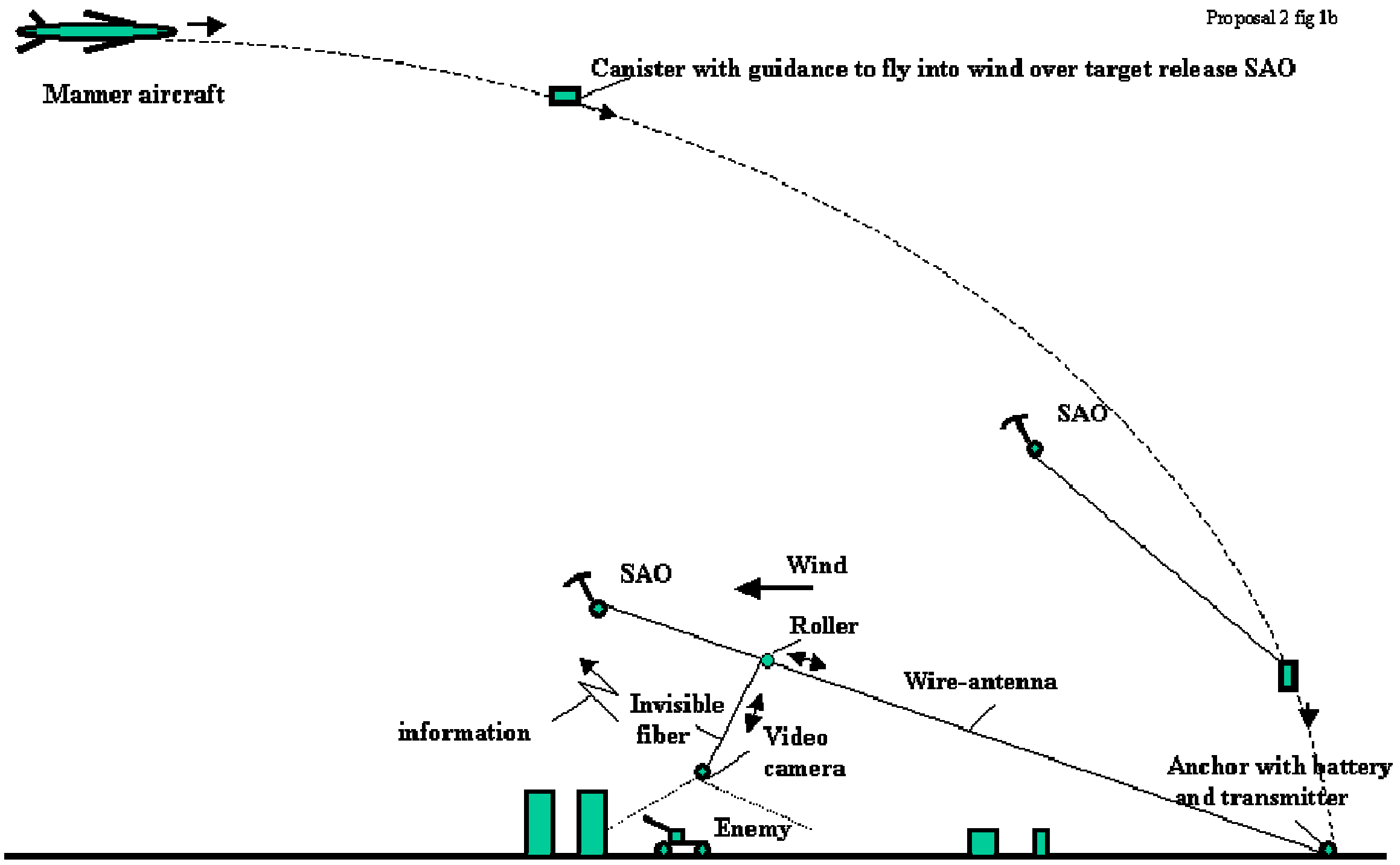

**Fig. 1**. Launching of MAO (SAO) from aircraft.

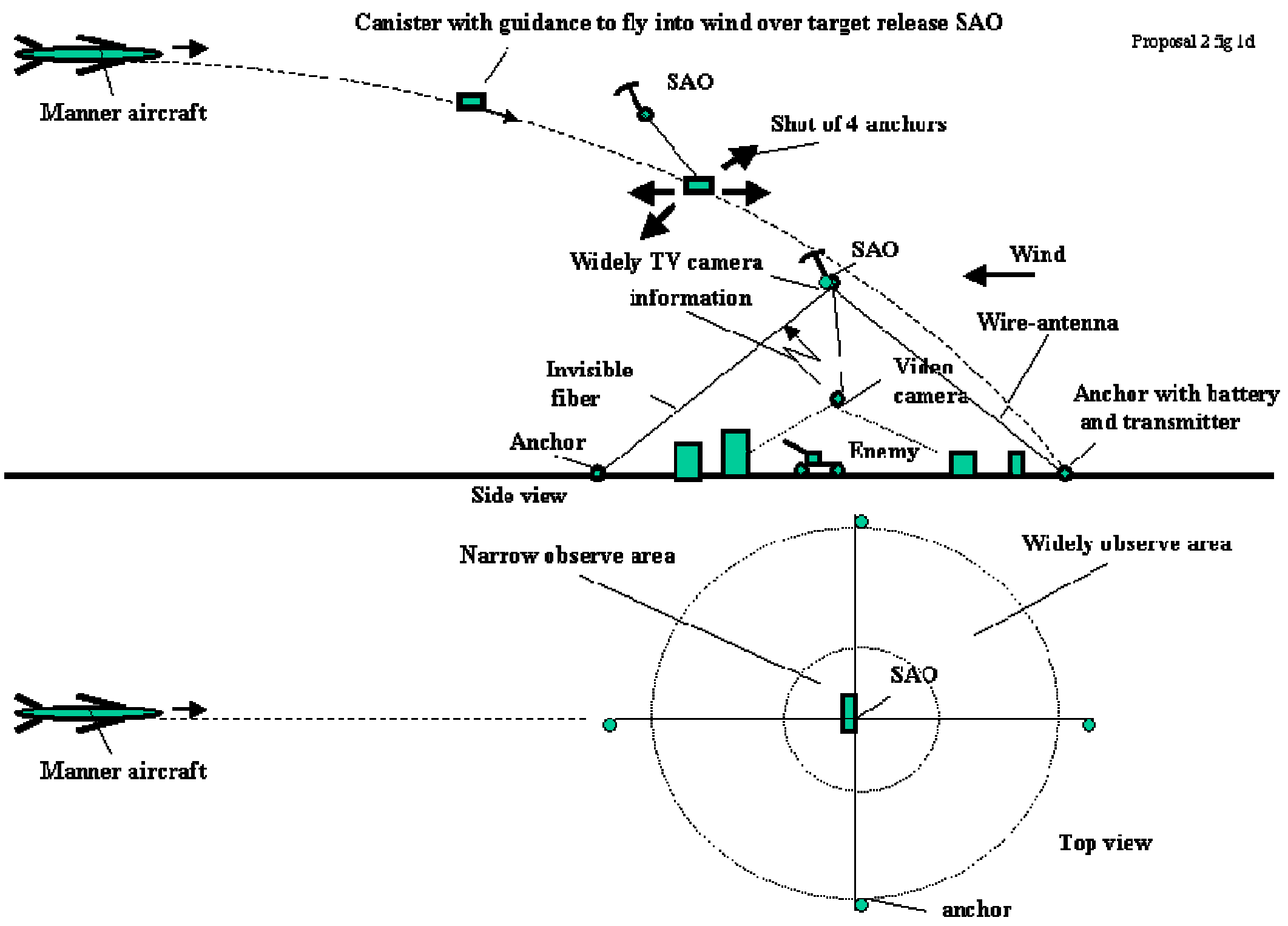

**Fig. 2.** Launching of four anchor fixed SAO from aircraft.

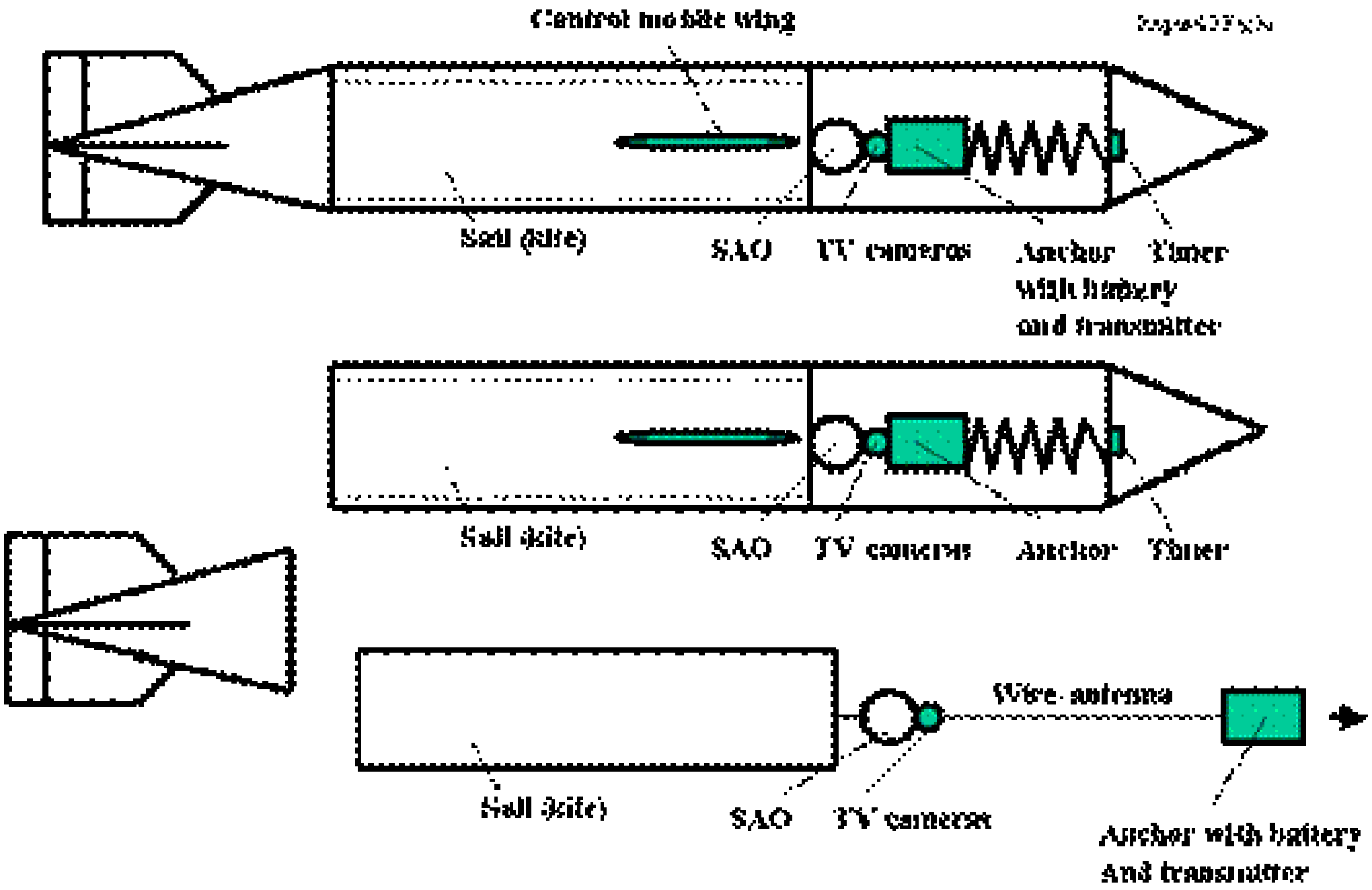

**Fig. 3**. Canister for MAO (SAO).

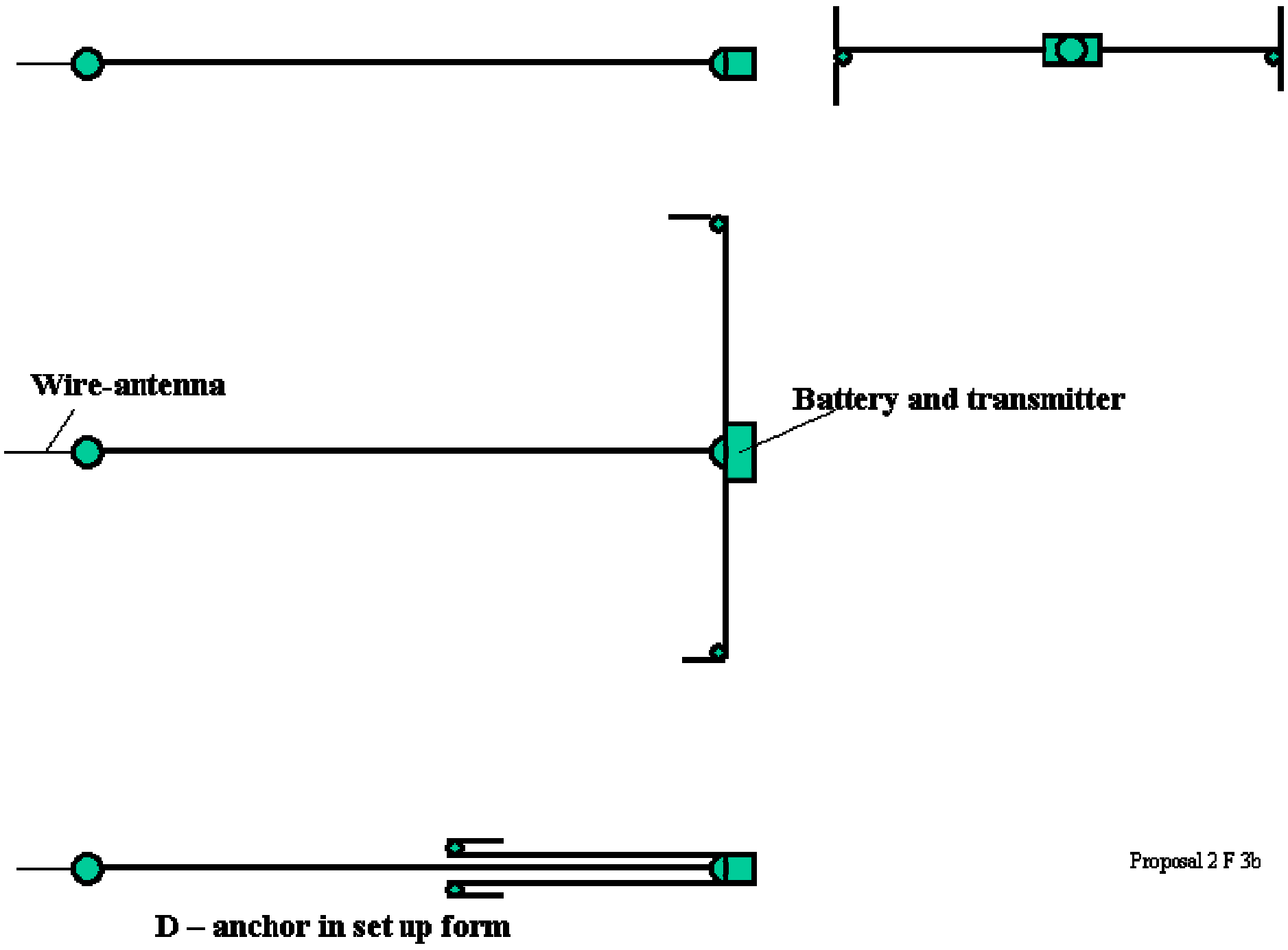

**Fig.4.** Anchor. D – anchor in set up form

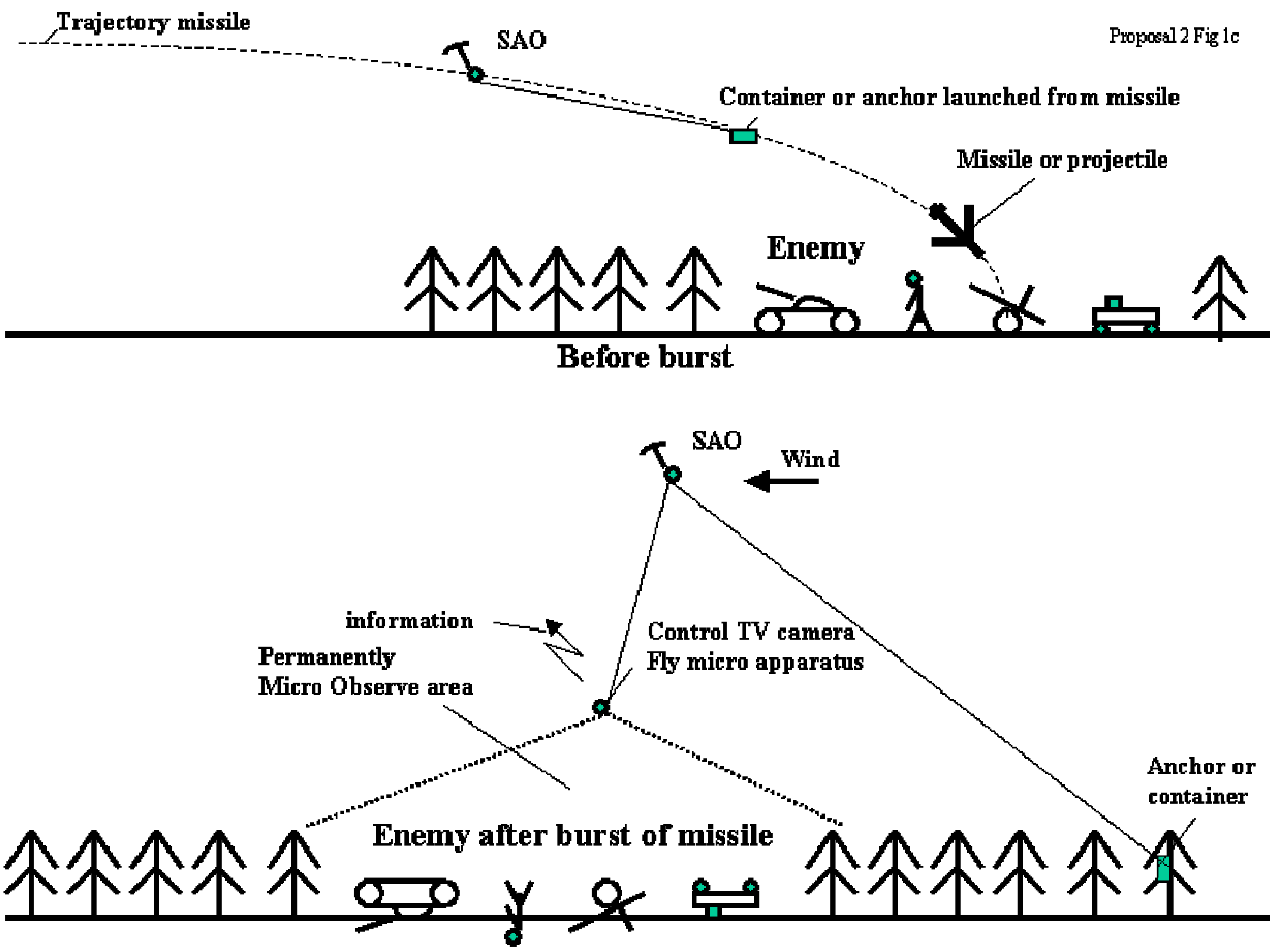

**Fig.5**. Launching of MAO (SAO) from missile or projectile.

b) An artillery shell launches the MAO. When the shell flies at an altitude 100-200 meters, the support device (parachute) is opened and the MAO is braked (fig.6). When the MAO reaches the ground, it is would likely be self-destroyed by a flight termination unit.

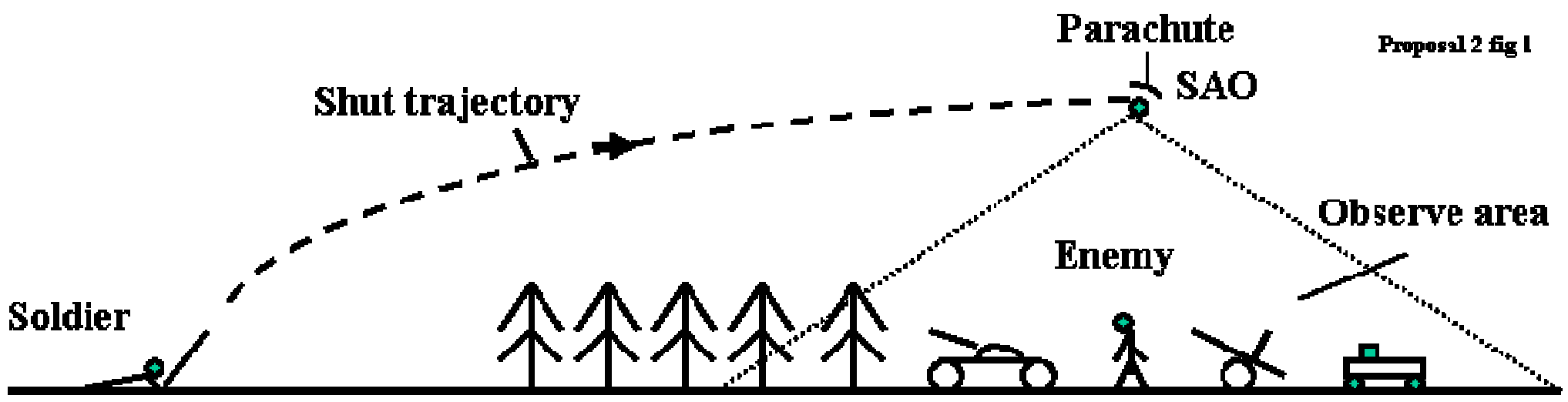

**Fig.6**. Suspended Aerial Observer MAO (SAO).

c) If prevailing wind blows towards enemy. The MAO may be operated as a kite as long as the wind speed remains never drops below some threshold value for more than some determinable period of time.

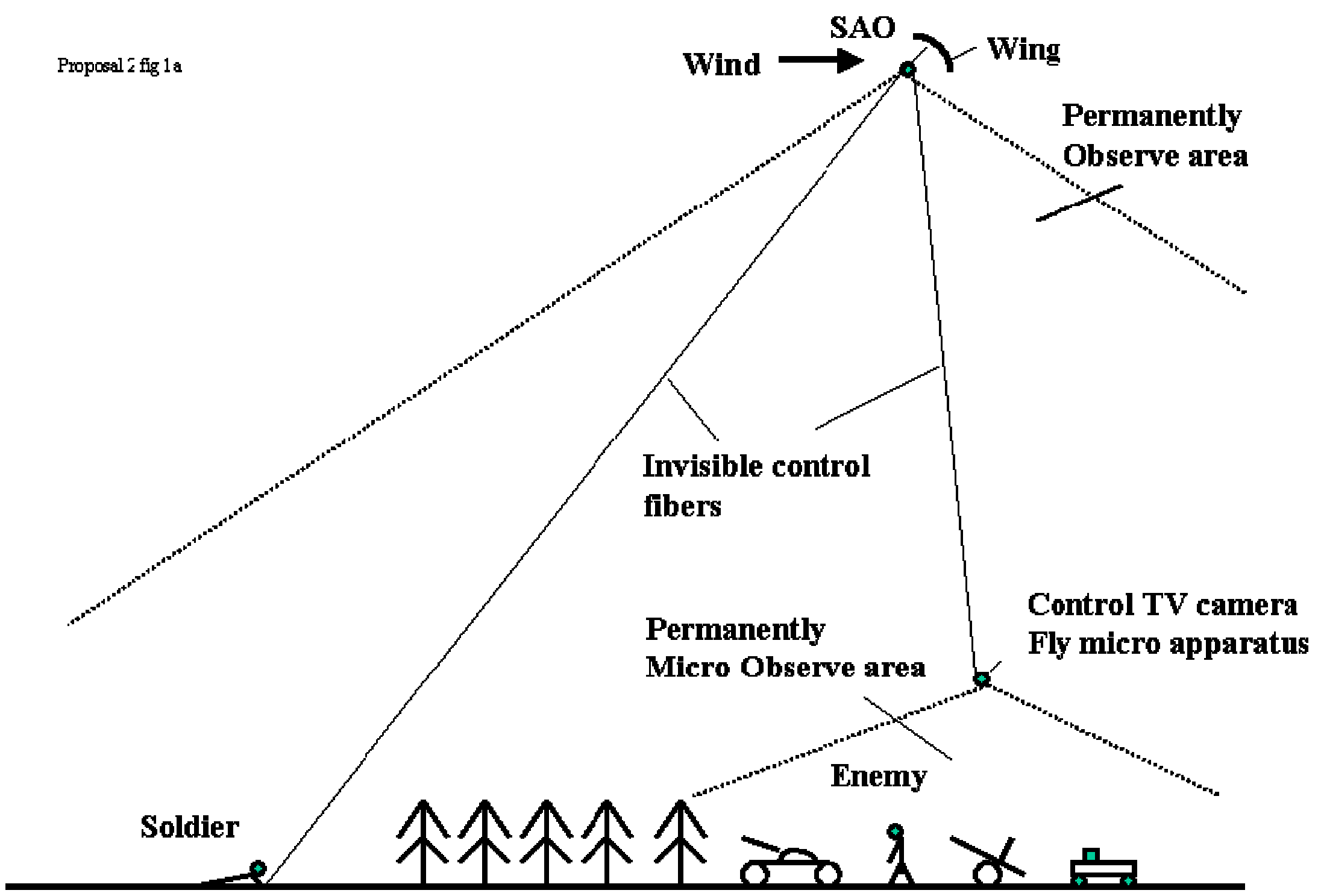

**Fig.7.** Launching Micro Air Eye (MAE) as kite.

d) If prevailing wind blows towards operator. The operator deploys the MAO in the direction of the desired site. At the apogee of the trajectory the brake parachute is opened, the MAO, connected to the shell by fiber, is slowed. The shell continues on its trajectory and falls behind the enemy location. The shell is used as anchor for the MAO (fig.8) which flies as a kite, lowers the camera, and permanently observes the enemy location.

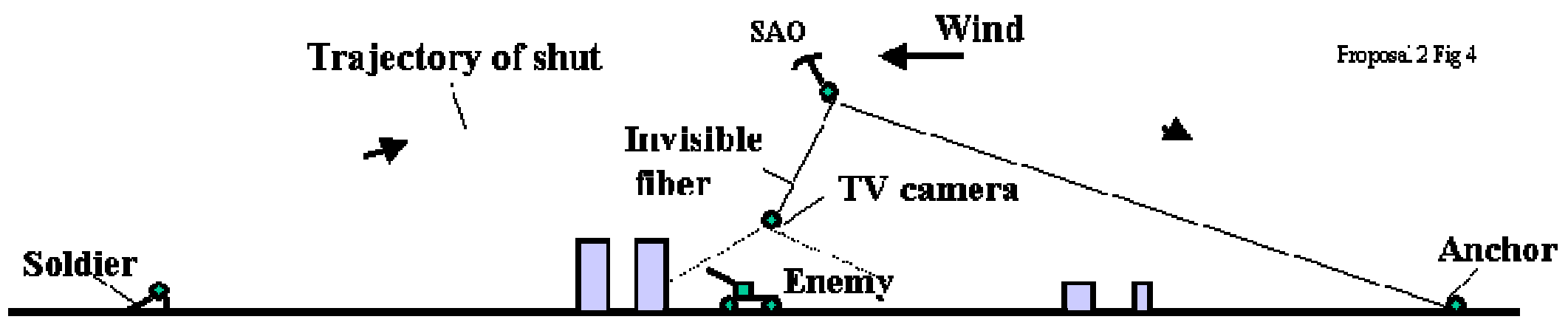

**Fig. 8.** Using MAO (SAO) when wind is from enemy.

e) If prevailing wind is orthogonal to line of sight between operator and observation site. The MAO is deployed in an orthogonal direction (fig.9) and is configured as in (d).

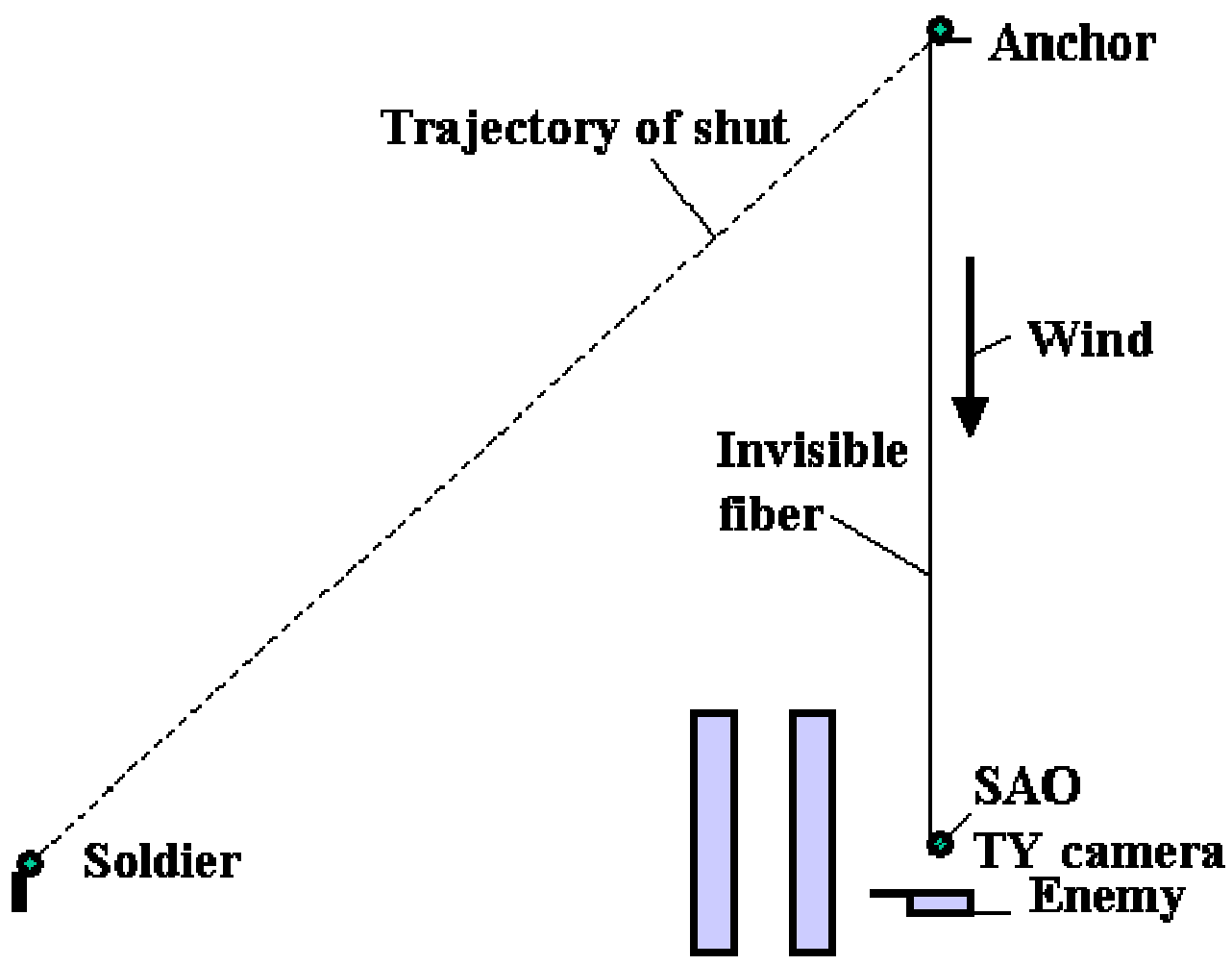

**Fig. 9.** Using MAO (SAO) when wind is from side (top view).

f) The camera fiber may be connected to the kite assembly via a controlled roller (fig.10). It is moved along the main cable and changes position and altitude. The MAO and/or camera assembly may also have a small vertical wing to increase stability and to allow maneuver for increasing the observable area (fig.11).

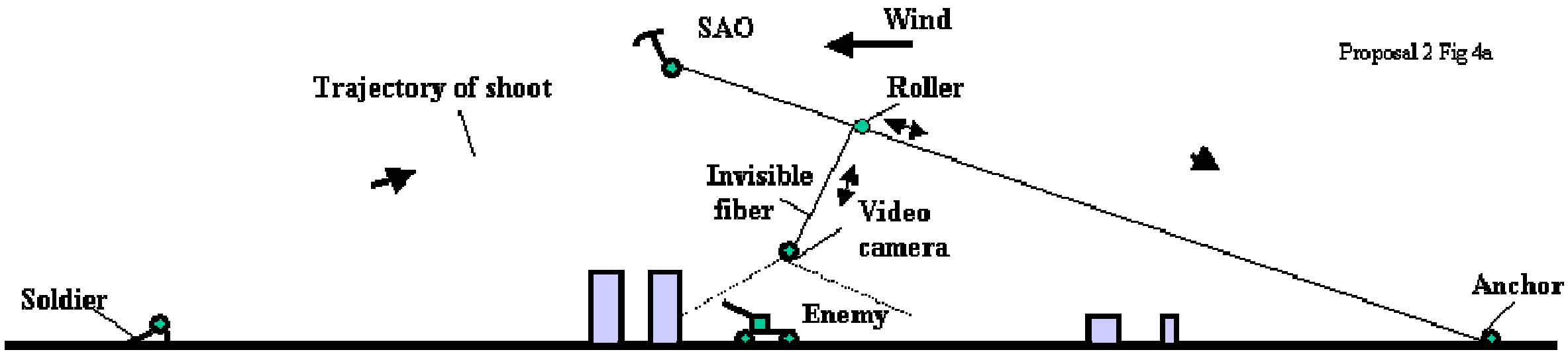

**Fig.10.** MAO with mobile TV camera.

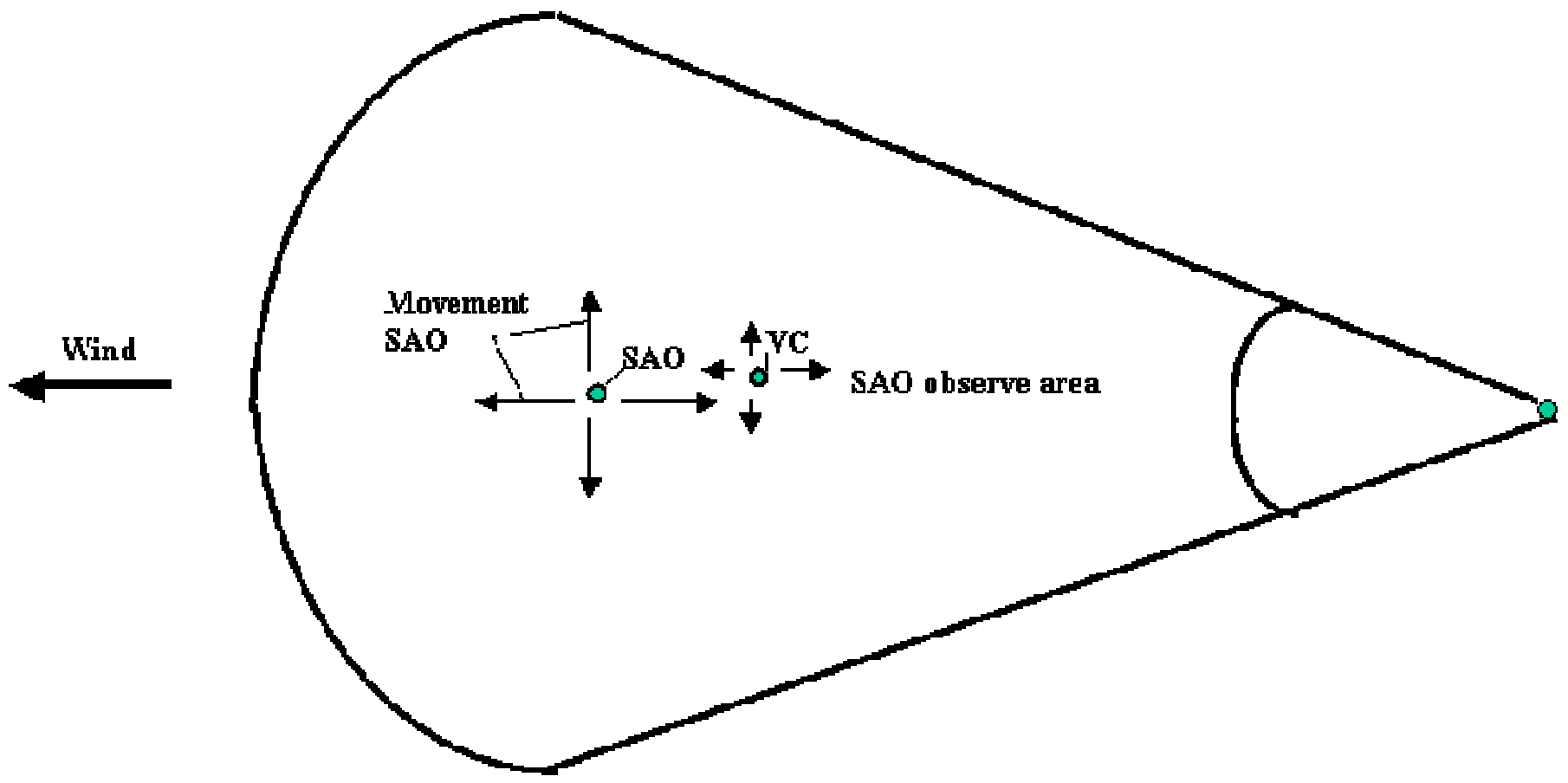

**Fig.11**. MAO observe area.

g) The MAO may be also suspended by an air balloon (diameter 1-1,5 foot, 25-40 cm)(fig.12). In this case, it may operate in windless weather. Of course the balloon would be easy to observe.

h)

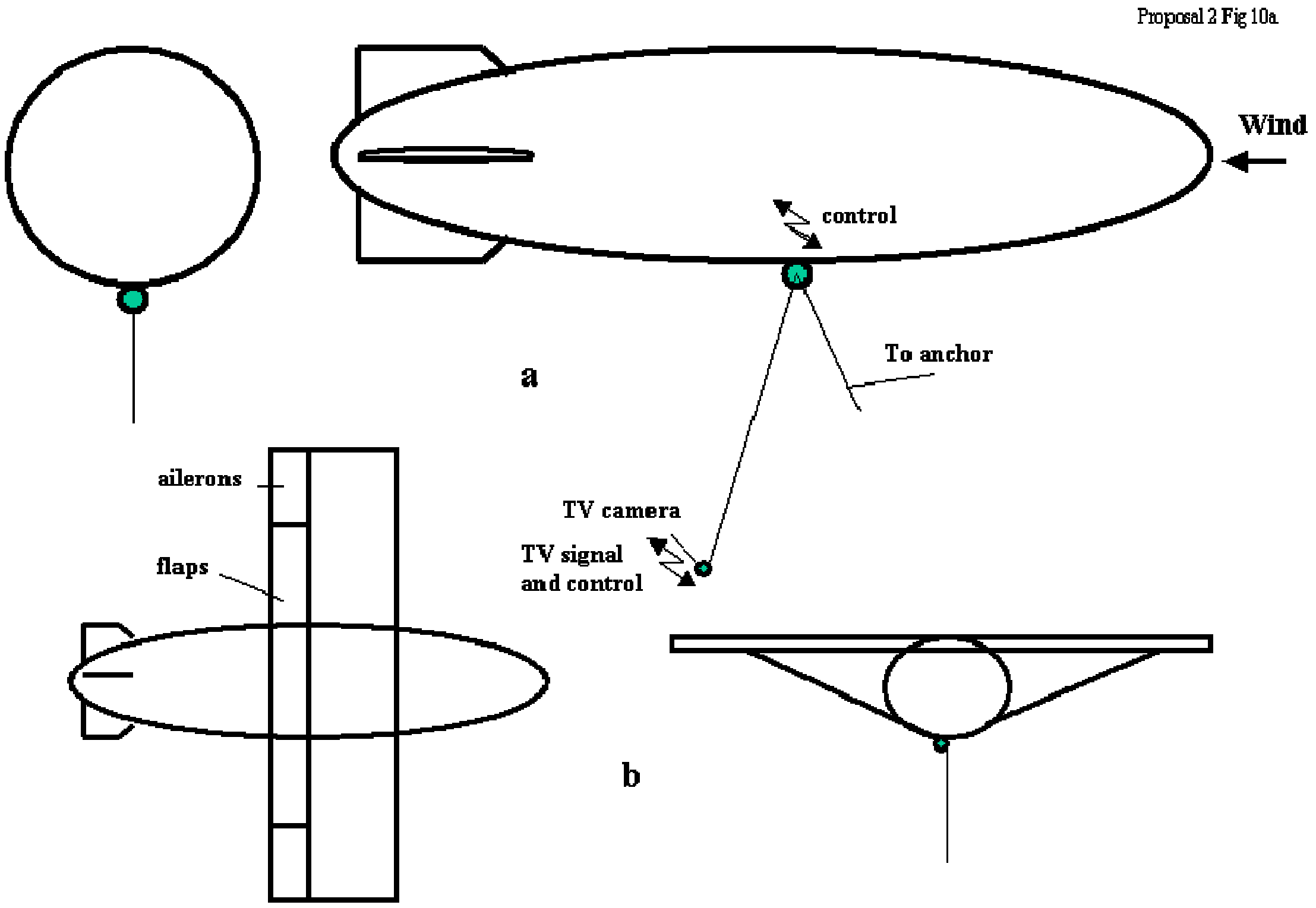

**Fig. 12**. Support balloon: a) without wing; b) with wing.

i) For more efficiency or to permanently observe a large area, the MAO may be connected to three to eight anchors (figs. 13-14). In this case the MAO position does not depend on wind direction.

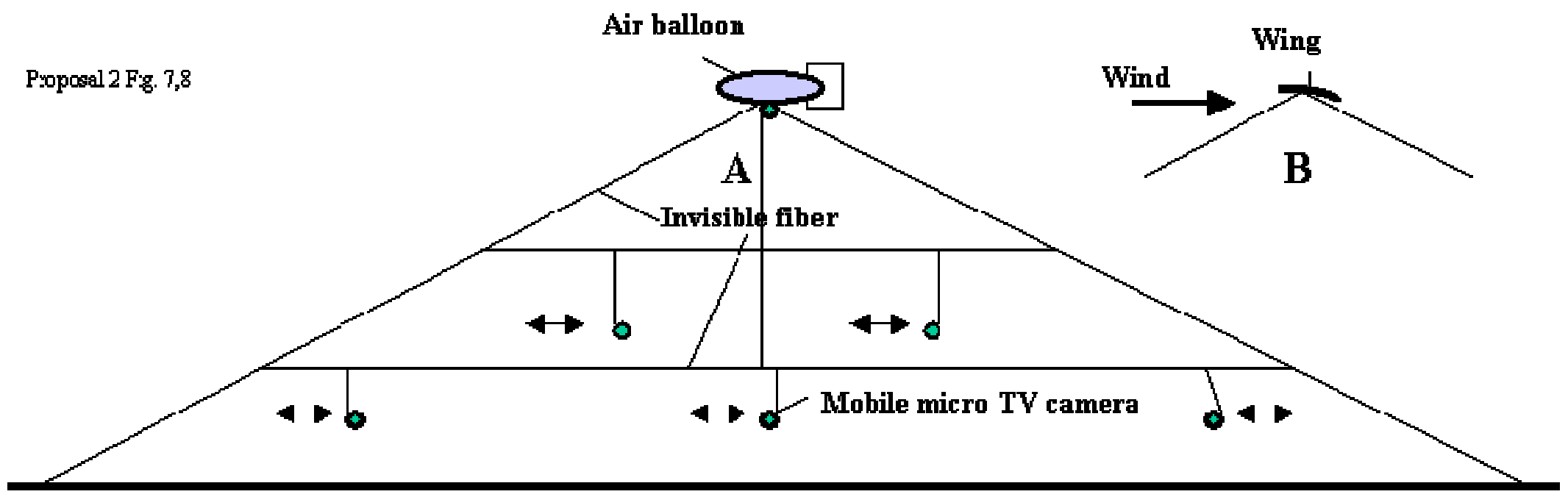

**Fig. 13**. Installation for Stationary observe area: (A ) with balloon, (B) with wing.

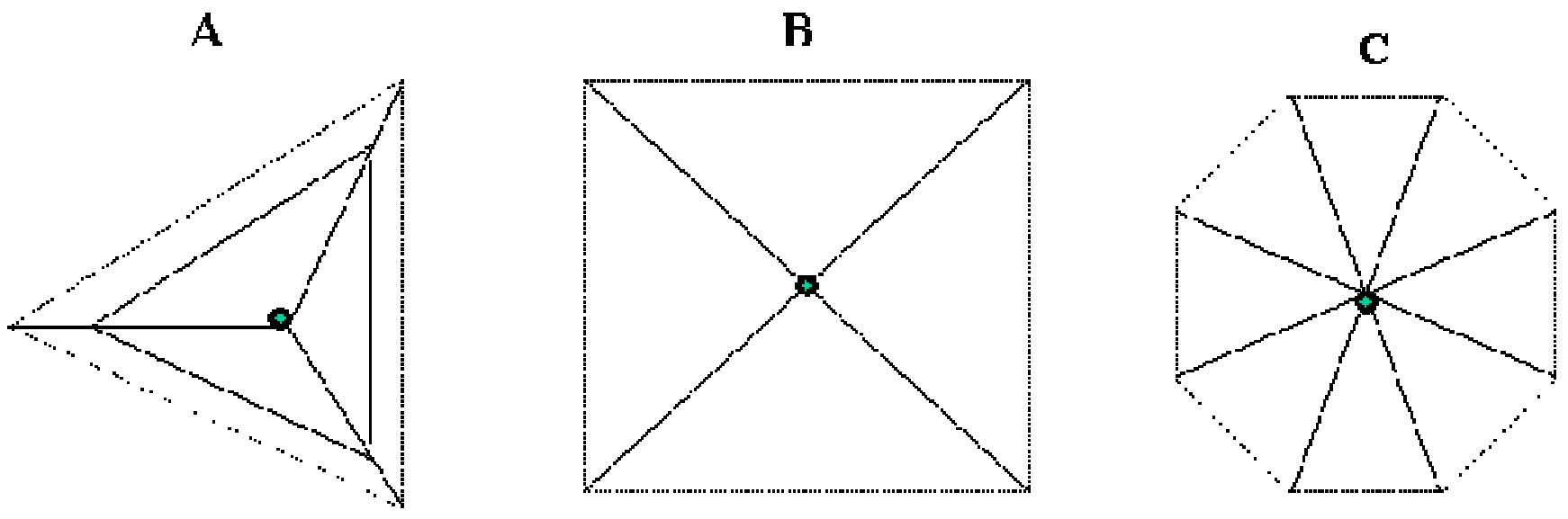

**Fig. 14.** Top view of the Installation of Fig.7. A – three conventional cables, B – four conventional cables, C – eight cables/

The support device or kite may have a sail form (Fig.15) or conventional airplane form (it is most efficient, fig.16). It must have control surfaces to control the MAO. The camera assembly itself may have a small vertical wing and flaps (figs.17), which allows it to move side to side and back and forth. The spool also allows changing the camera altitude and the flaps allow the operator to change the fiber angle.

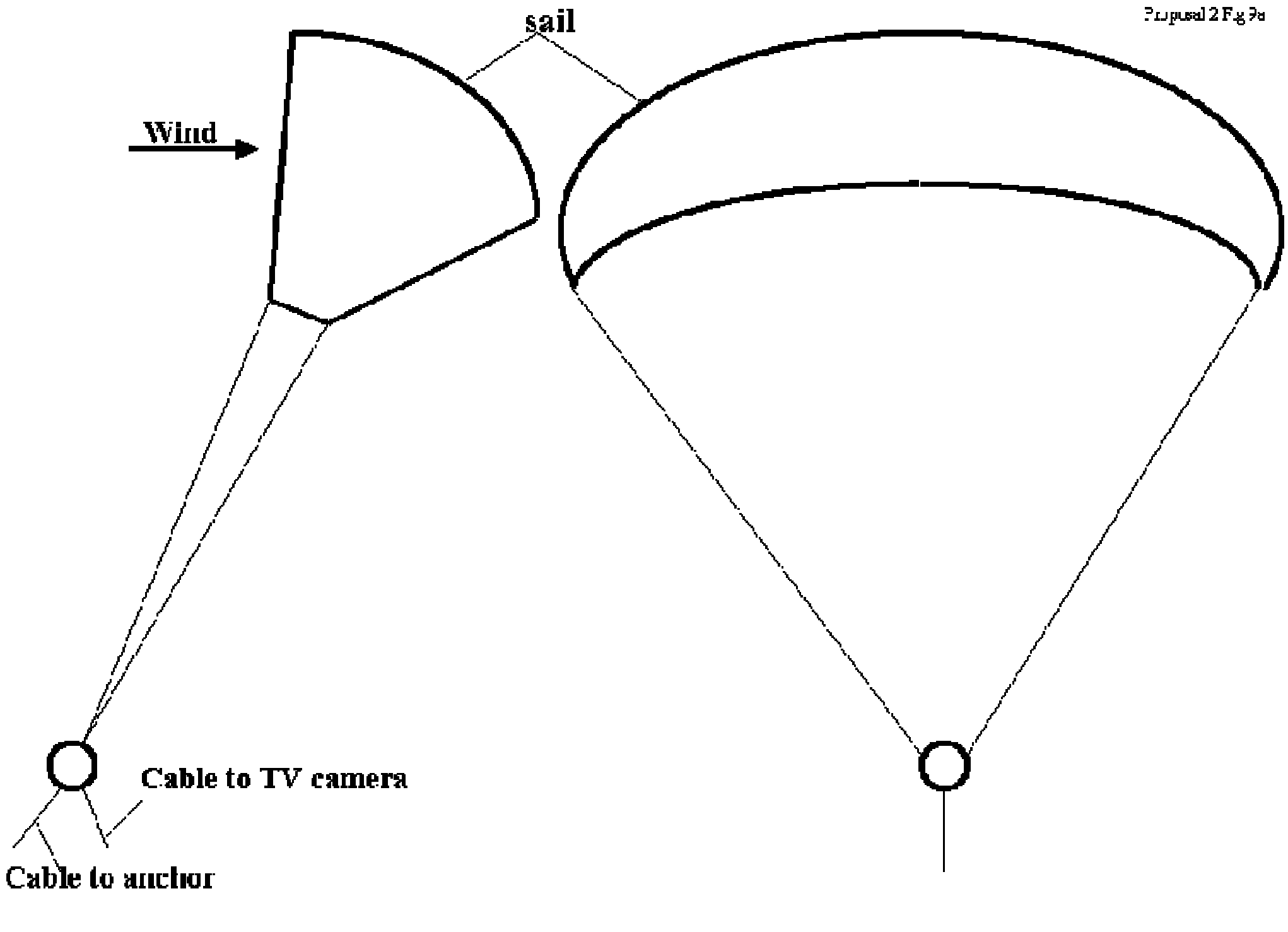

**Fig.15.** Support sail of MAO.

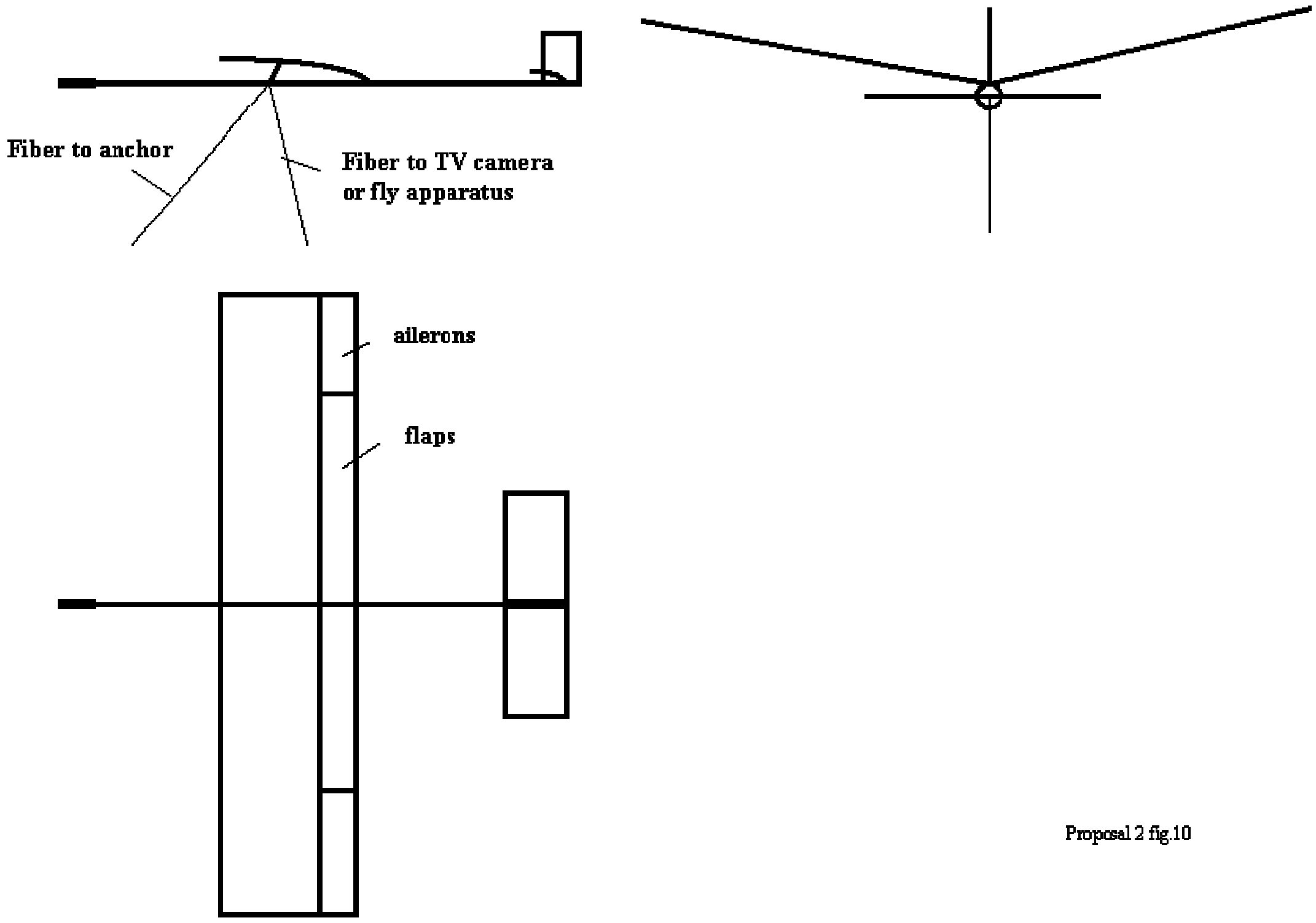

**Fig.16.** Base wing.

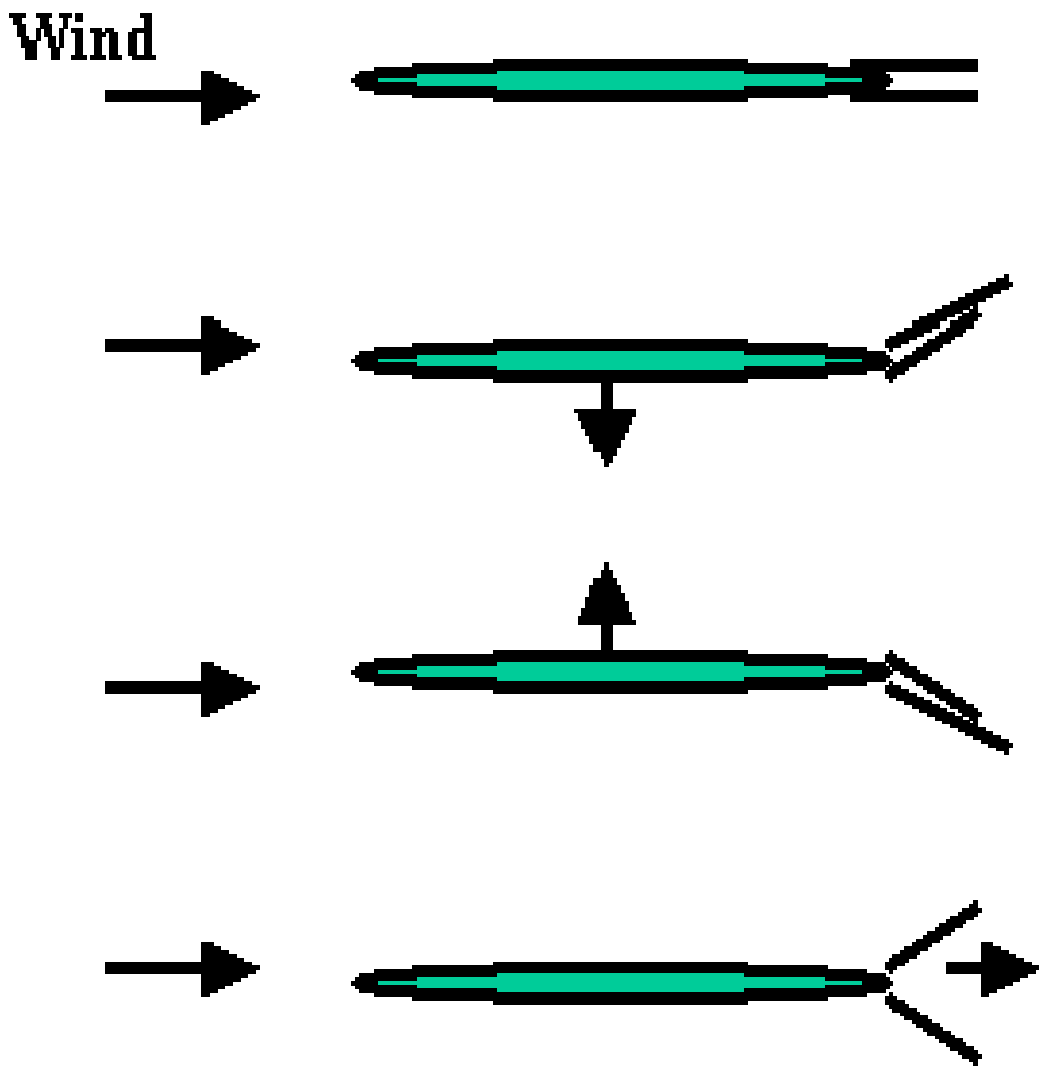

**Fig.17**. Control position of TV camera.

The closed loop stabilization and pointing of the camera assembly will require a gyro wind propeller (figs. 18-19). The high revolution wind propeller has loads at the blade ends (fig.19). One has a rigid connection to the camera station, and a swivel-spring connection with the vertical wing. The gyro propeller has a gyroscopic effect, which does not allow a sharp turn of the video cameras.

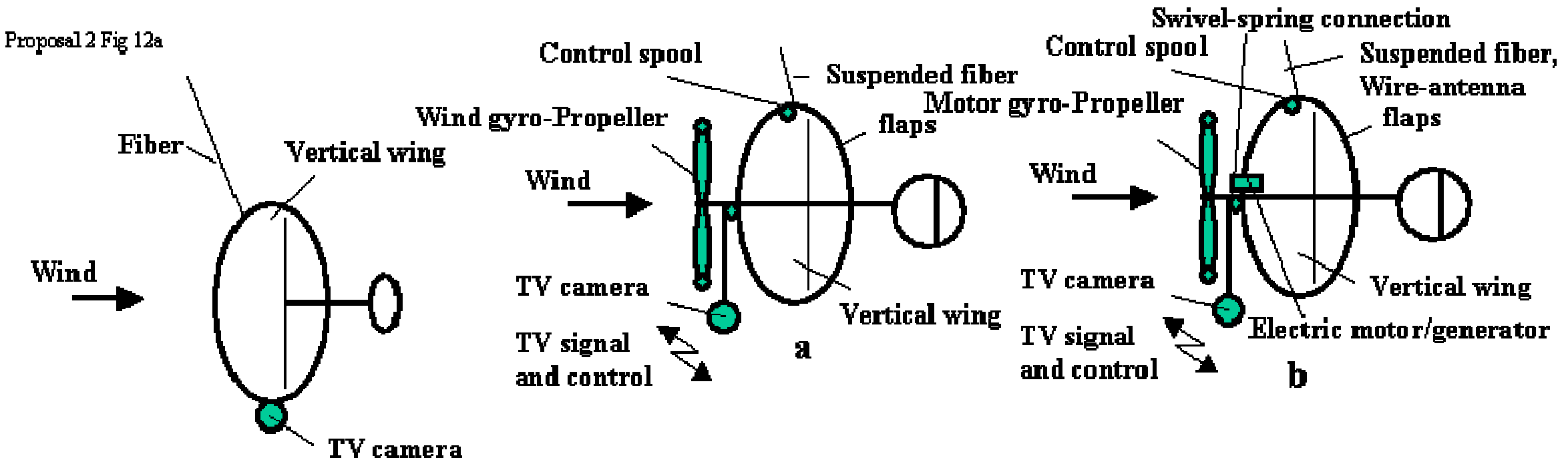

**Fig. 18**. Fly apparatus with wind or electric gyro and stabilized TY camera.

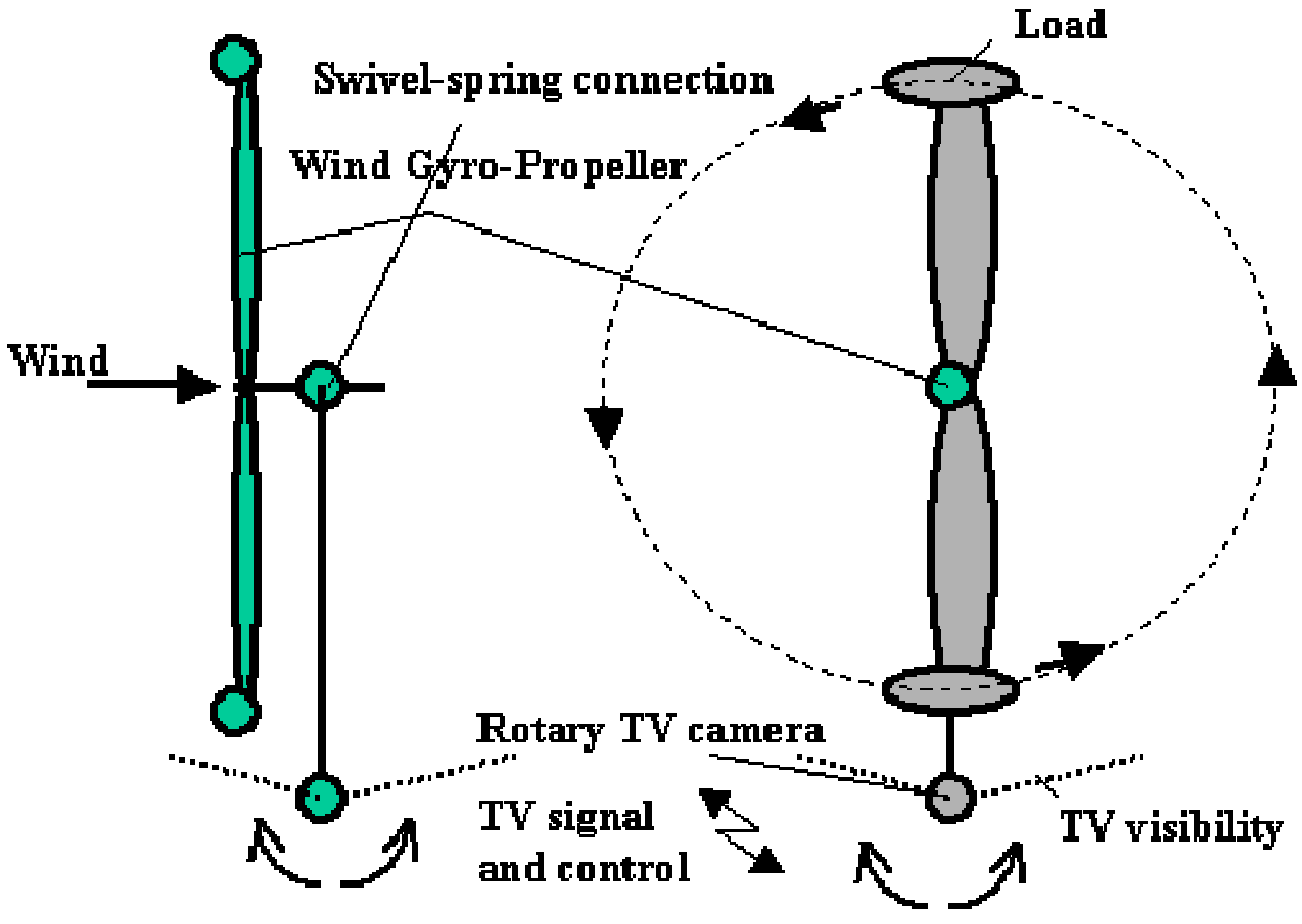

**Fig. 19**. Wind gyro stabilizer.

The camera assembly may have the two video cameras (fig.20): a widely angle fixed lens and a narrow angle swinging mobile camera. The widely angle camera allows the operator to observe a general picture, the narrow angle camera allows the operator to observe a small selected object.

If the wind at low altitude is small or absent, the camera assembly may have a small electric motor which rotates a gyro propeller (fig.20) which move the camera. The propeller (fig.18b) can also rotate a small electric generator to power-up the electronics.

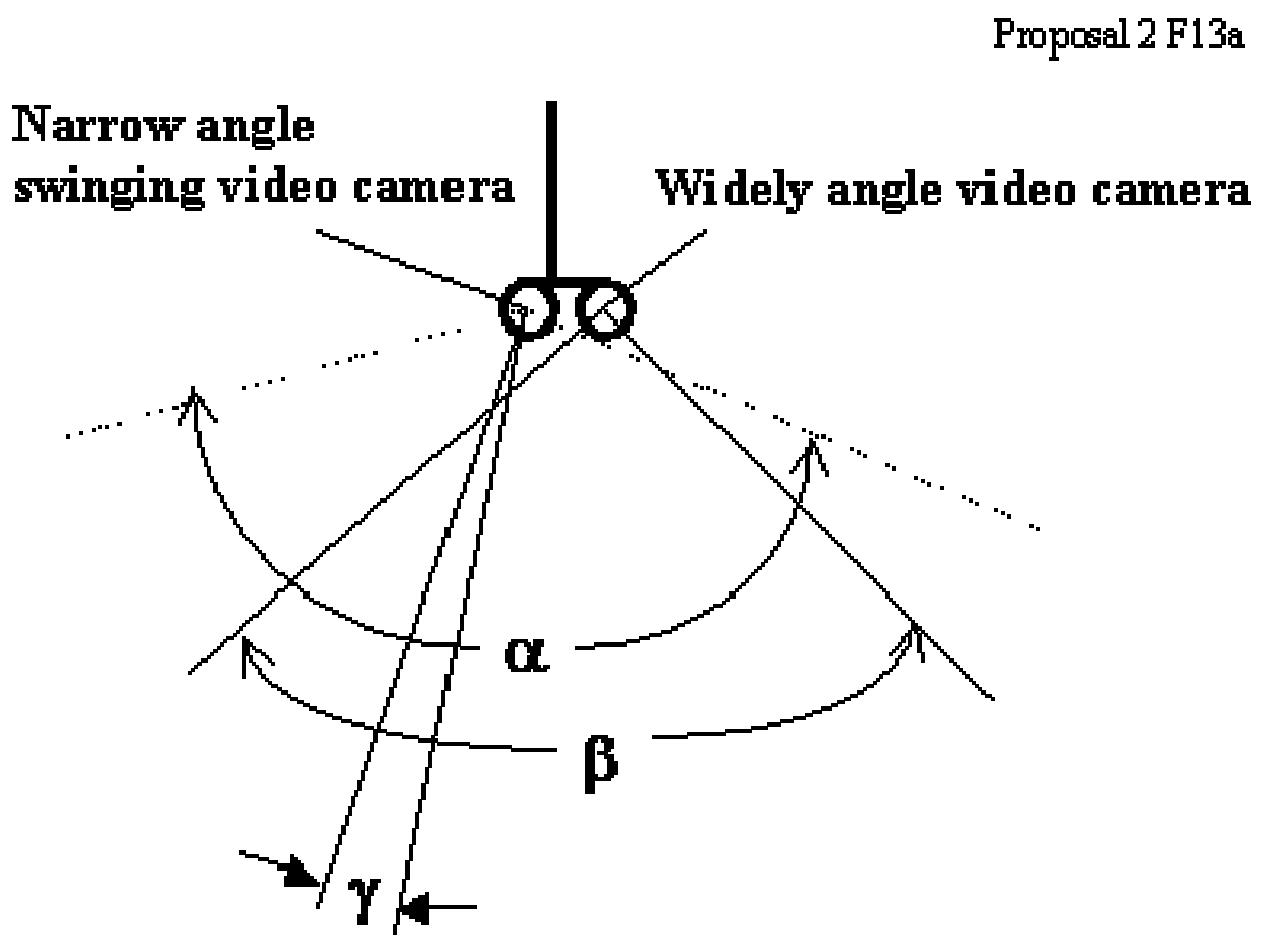

**Fig. 20**. Two cameras stabilized video station.

An auto-gyro propeller (diameter35-75 cm) (fig. 21) can be used as a support device. It may also be rotated by its motor or a ground driver (motor) through a cable transfer (fig.22).

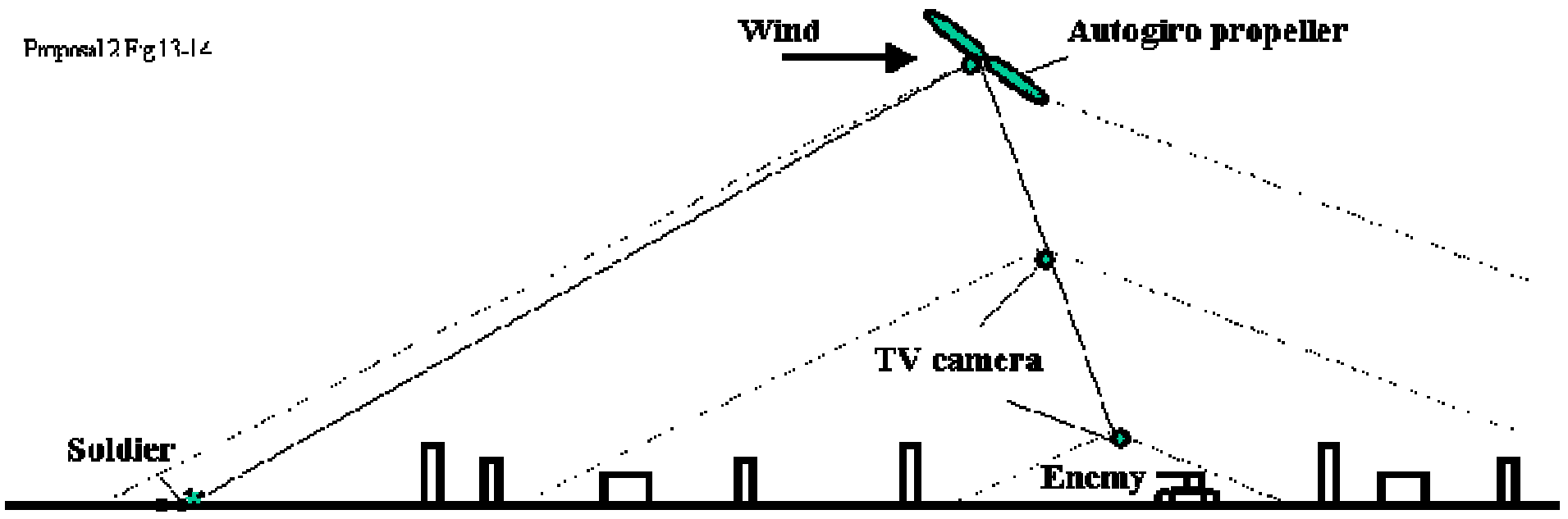

**Fig. 21**. Suspended Aerial Observer with passive autogiro propeller.

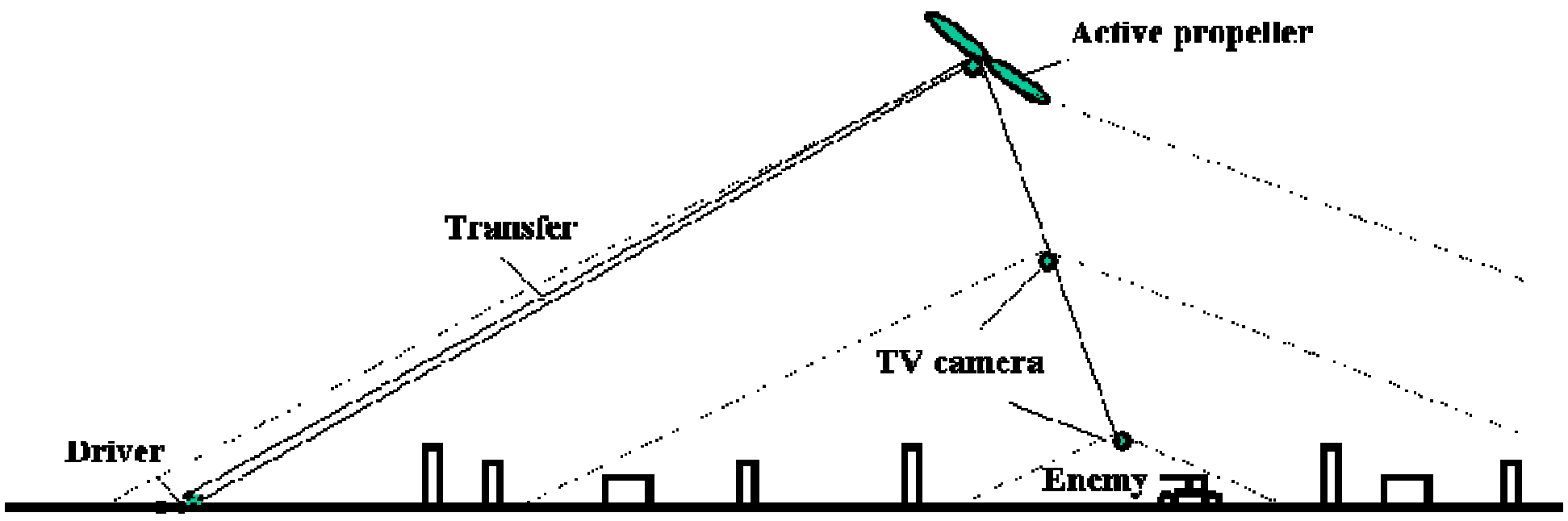

**Fig. 22**. Suspended Aerial Observer with active propeller.

The last variants are presented in figs. 23-26. In fig.23, the MAO uses an air balloon, a mobile controlled suspended video camera and a control suspended fiber. The air balloon (size 40-120 cm) is made from glass thin film, one is located at high altitude (200-500 m) and it only slightly visible from the surface. Fig.24 is the same as fig.23, but the MAO uses the propeller and motor as the support device. Figs. 25-26 show the control support device being used to increase the observed area.

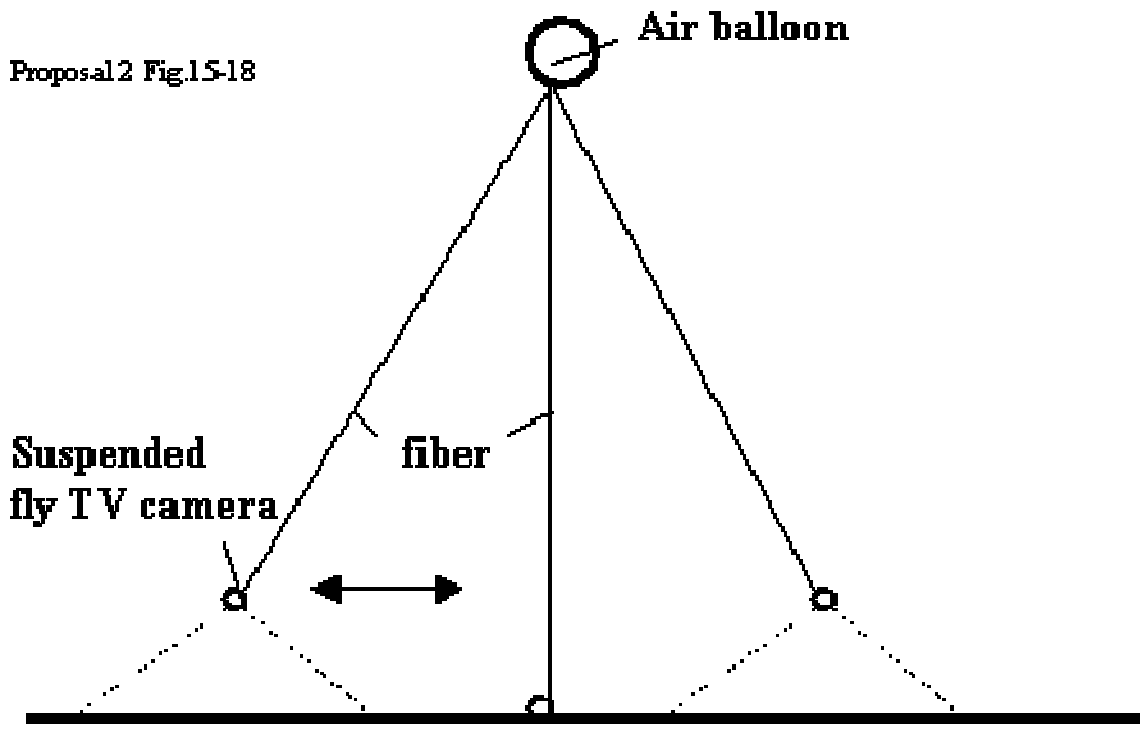

Fig. 23. Using of balloon MAO when no wind.

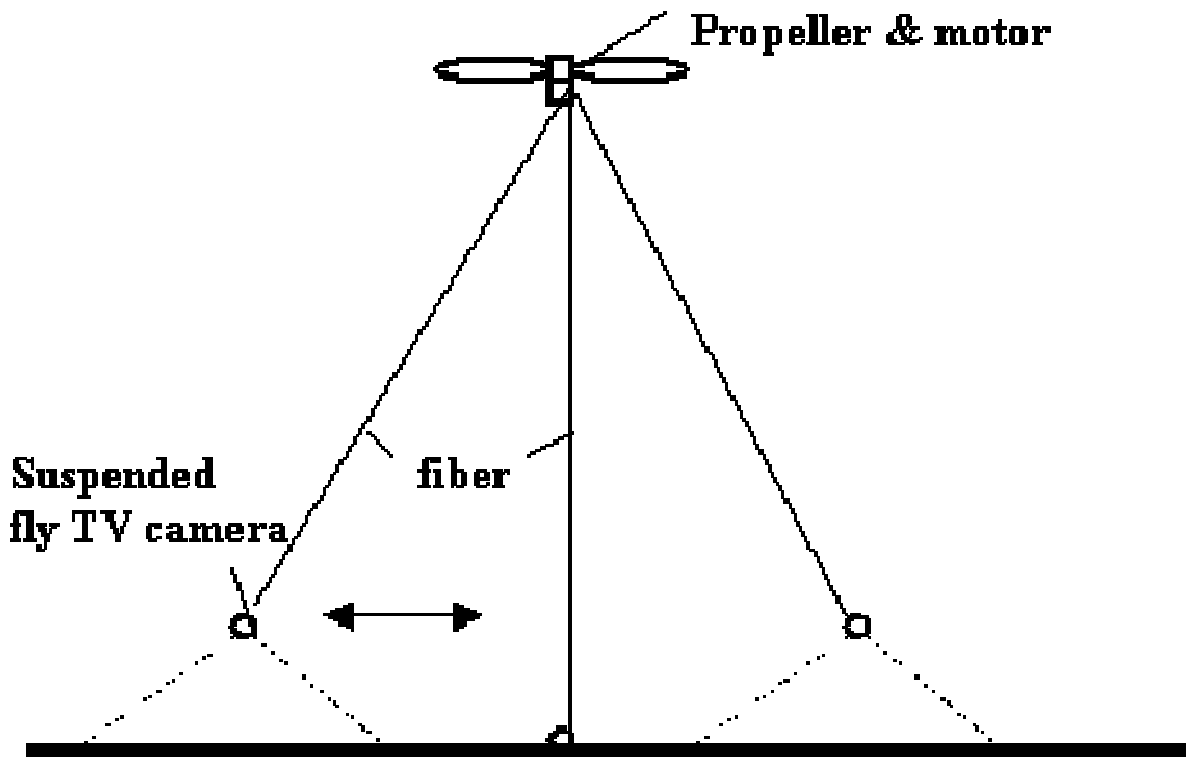

Fig. 24. Using of propeller MAO when no wind.

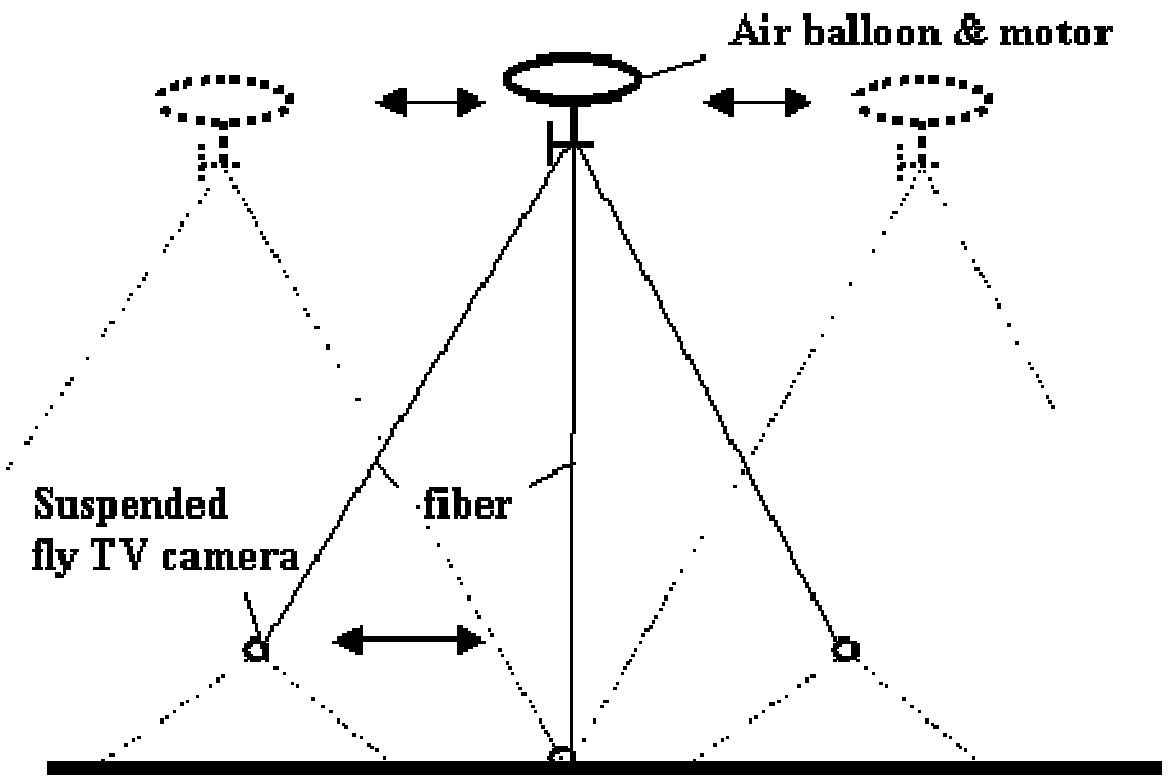

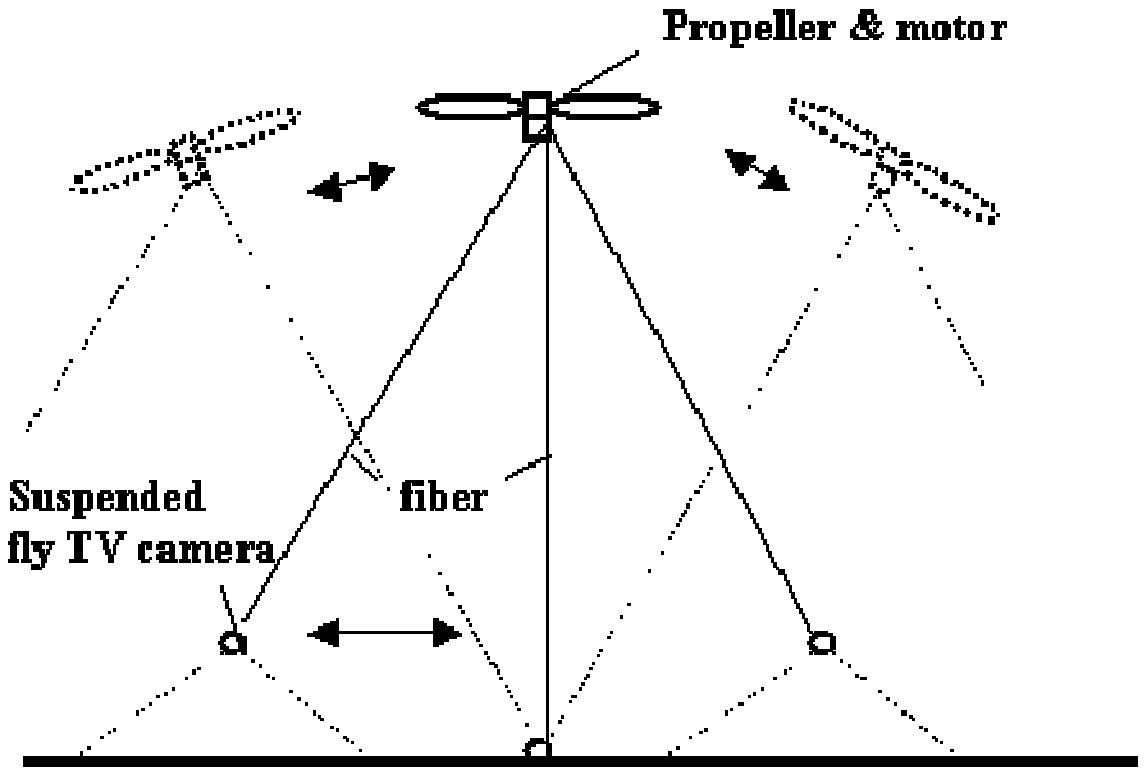

**Fig. 25**. Using of control balloon MAO when no wind. **Fig. 26**. Using cont.propeller MAO when no wind.

We can also use a small mobile controlled dirigible or helicopter. Our innovation is a connection to them by thin fiber a small mobile controlled video camera (and microphone) and lower them to an observe area or an enemy location. The main apparatus flies at high altitude; the video camera suspends at low altitude and permanently observes a needed area.

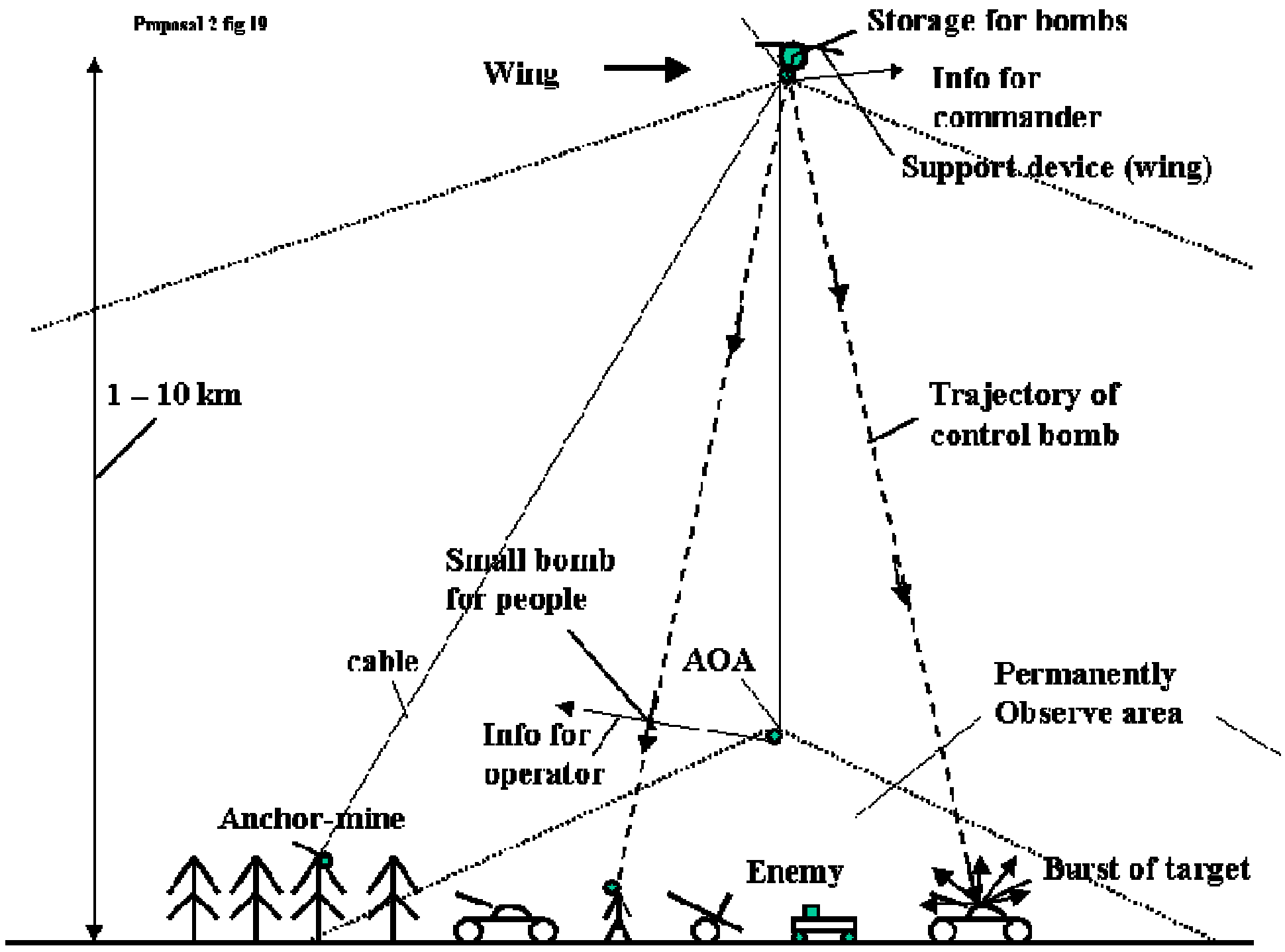

**Fig.27**. Air Observer and Annihilator (AOA).

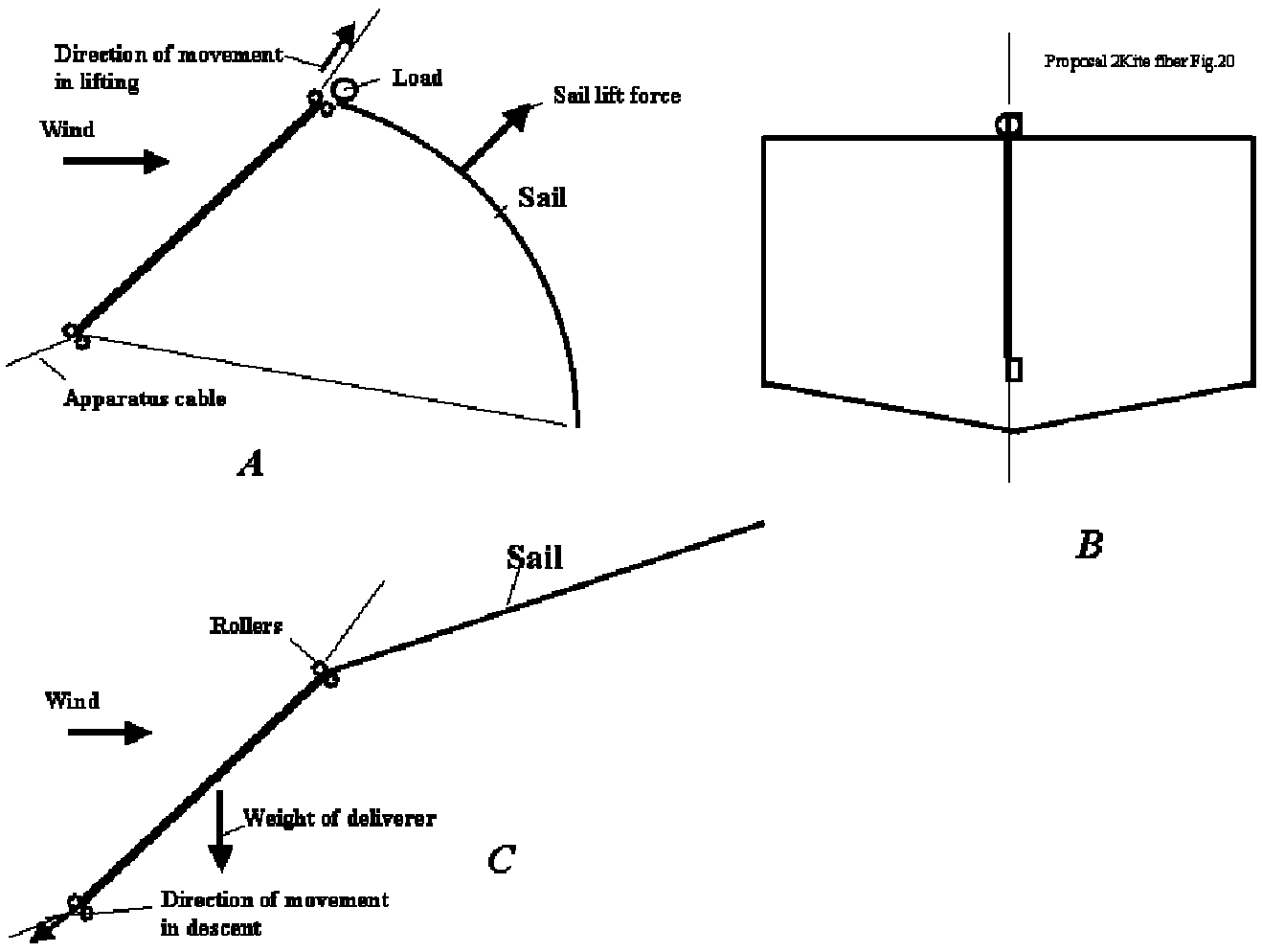

**Fig.28.** Load deliverer to AOA. ***A***- side view; B-front view; ***C***- Descent of Louder.

## 1.3. Advantages of Micro Air Observer in comparison with Micro Air Vehicle.

1. The suggested method may be applied in wind and weather when the MAV cannot be used because the wind is strong, turbulence, has scuds and twirls from building and trees. Indeed, to some extent, the greater the wind the better the performance of the MAO. Examining worldwide wind profiles, the probability of being able to use the MAO is greater than the probability of using a MAV. For example, for the average annual wind speed 6 m/sec the probability of MAO launching (V minimum 3 m/sec) is 0.85 when the launch is made from ground (from aircraft it is more 0.9); for MAV with maximum admissible speed 8 m/sec, the probability is only 0.65.

2. The offered method allows permanently observing a large area (some sq. km) a long time (up to months). The current MAV can observe about 0.1 sq. km a short time (2-5 min).
3. The offered method has power located on the ground and it can transmits the video and control signals a long distance (up to hundreds km). The MAO or MAO can have a small electric generator, which is rotated by the wind propeller, and thus the devices (video camera and control) can operate a long time.
4. The suggested method allows permanently observing inside building rooms through the building windows.
5. Apparatus may be deployed quickly.
6. The MAO control is simple and does not require special training to control - unlike a micro air vehicle. Fitting a MAV with an autopilot is difficult and when completed still requires the MAV to carry along additional power to persevere.
7. Both the MAO and MAV may be deployed from aircraft or artillery shells.
8. The MAO is simpler and less expensive than a MAV.

### 1.4. Lack of wind.

The MAO can only operate if the wind (at altitude) is more than some certain minimum (e.g. 2-3 m/sec). If the wind speed is less the minimum and MAO has special design, the MAO will land but can take off again when wind speed again becomes high enough to launch it. The MAO can also have a support air balloon. Special rotor kite can be supported by ground engine in windless weather.

Of course the logical approach is use the MAO to supplement the MAV and vice versa. The MAV and MAO have different conditions for application: the MAV in windless weather, MAO in wind weather.

### 1.5. Summary

The MAO is launched from an aircraft or is launched via an artillery shell or gun into an enemy observation area. As it enters a predetermined altitude 100-200 meters, the support devise is opened, the MAO is braked, the anchor (attached to MAO by fiber) is dropped down and connects the MAO to the ground (figs.1-10). The support device may be a small (less the one foot, 20-35 cm diameter) air balloon or small solid or inflatable kite or wing. The connection fiber also is invisible from a short distance (1-3 meter) because it may be fashioned very thin (as a hair) and is made from transparent and strong artificial material. If there is a small wind (at most places the wind is 80-90% days in year at ground, see a research below) the MAO will be supported at altitude by a small wing or wing sail (figs.5-9, 15). In other case, one can be supported a special MAO having a balloon with an inflatable kite or wing (fig.12-13). The support device (kite or wing) may be located at high altitude (1 or more km, different from MAO) and be connected to MAO by a thin fiber. There is sufficient wind at these high altitude about 95-98% of the year. Of course a major benefit is that the batteries can be located on the ground and so the MAO can operate for a long time. The MAO can also have a long antenna located at a high altitude and transmit a video and control signal for quite some distance.

The operator can observe the area of interest for a very long time (even weeks). If it is not needed in observation and no enemy, the soldier can reel it in up to the anchor (the anchor must have a radio-locator) and reel the thread (fiber) and get his device back. The MAO can be also launched as conventional kite for observe nearest closed area, especially, if the wind blows in enemy side.

The offered method and MAO has the following advantages in comparison of MAV:

1. No needs for developing a new top technology, which can confine Research, Development, and Design (as Micro engine, flight control, micro aerodynamics, autopilot, and navigation system).
2. The cost of Research and Development (R&D), design of MAO is less in 10 times that MAV.
3. The time of reconnaissance (observation) increases from 2-10 min to several weeks. The signals can be transmitted to long distance.

4. The using and control of MAO are simpler than MAV (not necessary in special training for soldiers or autopilot).
5. MAO can be used in wind and bad whether.
6. MAO uses the video devices, radio control, and communication developed for MAV.
7. MAO is cheaper then MAV and can be R&D and manufactured in short time.
8. MAO may be invisible for enemy.
9. There are a permanently air flow at high altitude

**Table 1**. The MAO data with comparison future MAV data:

| Parameter | MAO | best MAV |
|---|---|---|
| Size [cm] | 3 x 3 x 15 cm (in packet form). | 20 x 15 x 5 cm |
| Weight [grams] | 5-95 g | 5-95 g |
| Max range [km] | up 3-20 km (depend from start) | 1-2 km |
| Time observation | weeks (permanently) | 2-5 minutes |
| Area of observation | 1 square mile (permanently) | 0.02-0.05 sq. miles |
| Cost of production $ | same with MAV | same with MAO |
| Visibility | invisible | visible |

**1.6. Additional possibility for support MAO (SAO).**

High altitude wind has another important advantage. It is stable and constant. This is true practically everywhere.

Especially in the troposphere and stratosphere, the wind currents are powerful and permanent. For example, at an altitude of 5 km, the average wind speed is about 20 M/s, at altitude 10-12 km it may reach 40 m/s (at latitude of about 20-35$^0$N).

There are permanent jet streams at high altitude. For example, at $H$ = 12-13 km and about 25$^0$N latitude. The average wind speed at the it core is about 148 km/h. The most intensive portion, with a maximum speed 185 km/h latitude 22$^0$, and 151 km/h at latitude 35$^0$ in North America. On a given winter day speeds in the jet core may exceed 370 km/h for a distance of several hundred miles along the direction of the wind. Lateral wind shears in the direction normal to the jet stream may be 185 km/h per 556 km to right and 185 km/h per 185 km to the left.

Reference: *Science and Technolody,v.2, p.265.*

# 2. Theory and computation of MAO (SAO)

## 2.1. Wind

### (speed, duration, altitude distribution, speed distribution)

Wind is important element of the offered method. In MAV the wind is only obstacle which gives trouble for operator. The wind vortexes from buildings and trees are located at near Earth surface. One can set down to MAV to a ground. If wind is more then 6-8 m/sec, the flight of MAV can be impossible. In the MAO the wind is necessary for support of the apparatus. If wind is less a minimum (for example, 3 m/sec) the MAO, video camera lends to ground and can be take-off again when the wind will stronger. If the wind is very strong, the connection cable or a MAO wing can be damage.

We can calculate the minimum and maximum admissible wind for MAO and estimate it for MAV. Our purpose is estimation of time (% or a number of days in year) when the MAO and MAV can operate.

**Annual average wind speed.** On fig.2-1 is the accompanying map of the United States Annual Average Wing Speed taken from *Wind Energy Resource Atlas of the United States.* The map was published in 1987 by Battelle's Pacific Northwest Laboratory for the U.S. Department of Energy.

The complete atlas can obtained by writing the American Wind Energy Association or the National Technical Information Service. The same maps are around the world. They are presented in Attachment 6. The maps show the average wind speed at altitude 10 and 50 meters. This speed is 5-6 m/sec.

**Wind speed and Height**. Wind speed increases with height. The speed may be computed by equation

$$\frac{V}{V_0}=\left(\frac{H}{H_0}\right)^{\alpha}, \qquad (2\text{-}1)$$

where $V_0$ is the wind speed at the original height, $V$ the speed at the new height, $H_0$ the original height, $H$ the new height, and $\alpha$ the surface roughness exponent (Table 2-1).

**Table 2-1.** Typical Surface Roughness Exponents for Power Law method of Estimating Changes in Wind Speed with Height

| Terrain | Surface Roughness Exponent, α |
|---|---|
| Water or ice | 0.10 |
| Low grass or steppe | 0.14 |
| Rural with obstacles | 0.20 |
| Suburb and woodlands | 0.25 |

Reference: P.Gipe, Wind Energy comes of Age, 1995.

The result of computation of equation (2-1) for different α is presented at fig.2-2. The wind speed increases on 20-50% with height.

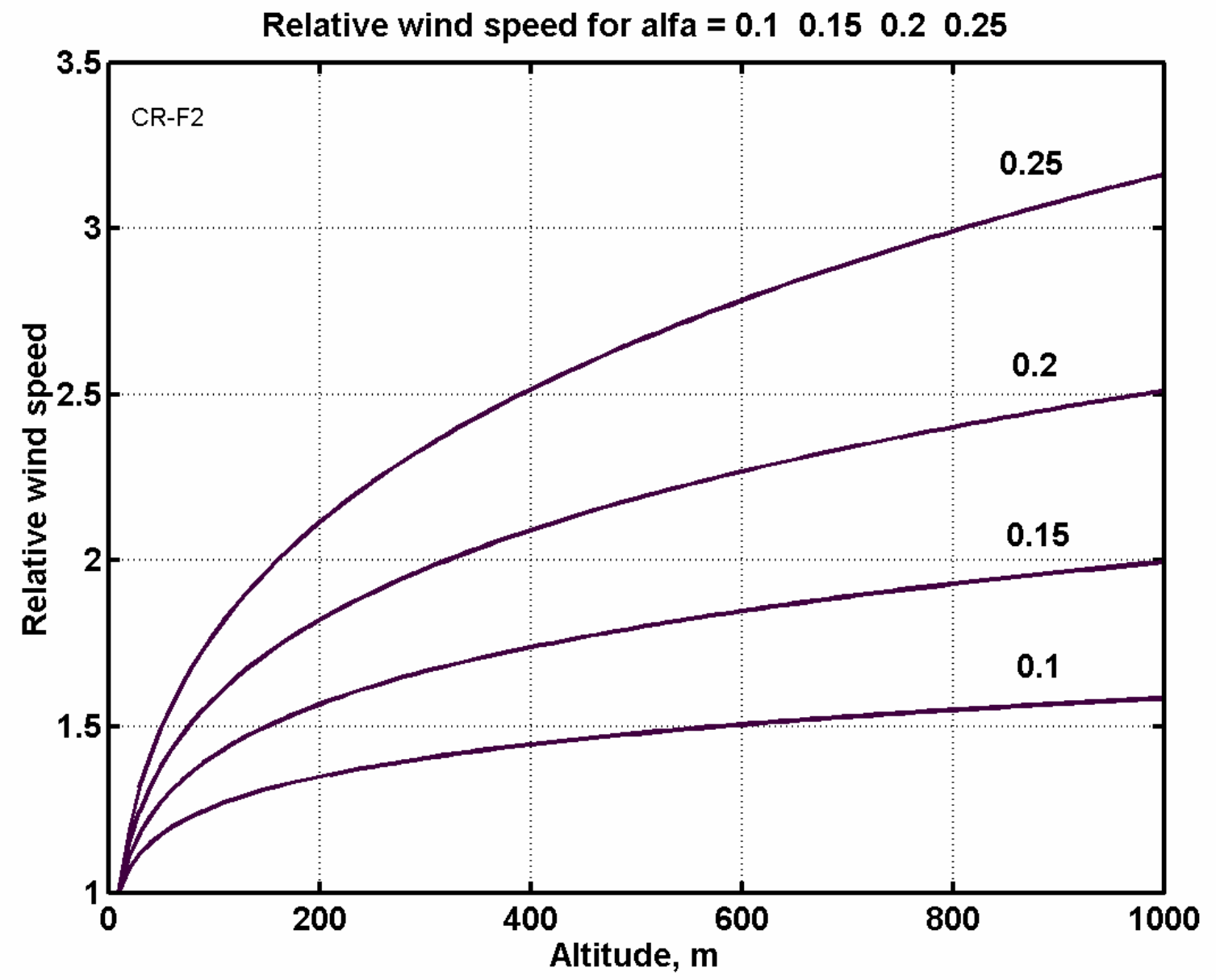

**Fig.2-2**. Relative wind speed via altitude and Earth surface. For sea and ice α = 0.1.

**Annual Wind speed distribution.** Annual speed distributions vary widely from one site to another, reflecting climatic and geographic conditions. Meteorologists have found that Weibull probability function best approximates the distribution of wind speeds over time at sites around the world where actual distributions of wind speeds are unavailable. The Rayleigh distribution is a special case of the Weibull function, requiring only the average speed to define the shape of the distribution.

Equation of Rayleigh distribution is

$$f_x(x) = \frac{x}{\alpha^2}\exp\left[-\frac{1}{2}\left(\frac{x}{\alpha}\right)^2\right], \quad x \geq 0, \quad E(X) = \sqrt{\frac{\pi}{2}}\alpha, \quad Var(X) = \left(2 - \frac{\pi}{2}\right)\alpha^2, \quad \text{(2-1a)}$$

where $\alpha$ is parameter.

Fig.2-3 presents the annual wind distribution of average speeds 4, 5, and 6 m/s. Table #2-2 gives Rayleigh Wind Speed Distribution for Annual Average Wind Speed in m/s. These data gives possibility to easy calculate the amount (percent) days (time) when MAO or MAV can operate in year (fig.2-3a). It is very important value for the estimation efficiency of offered devices.

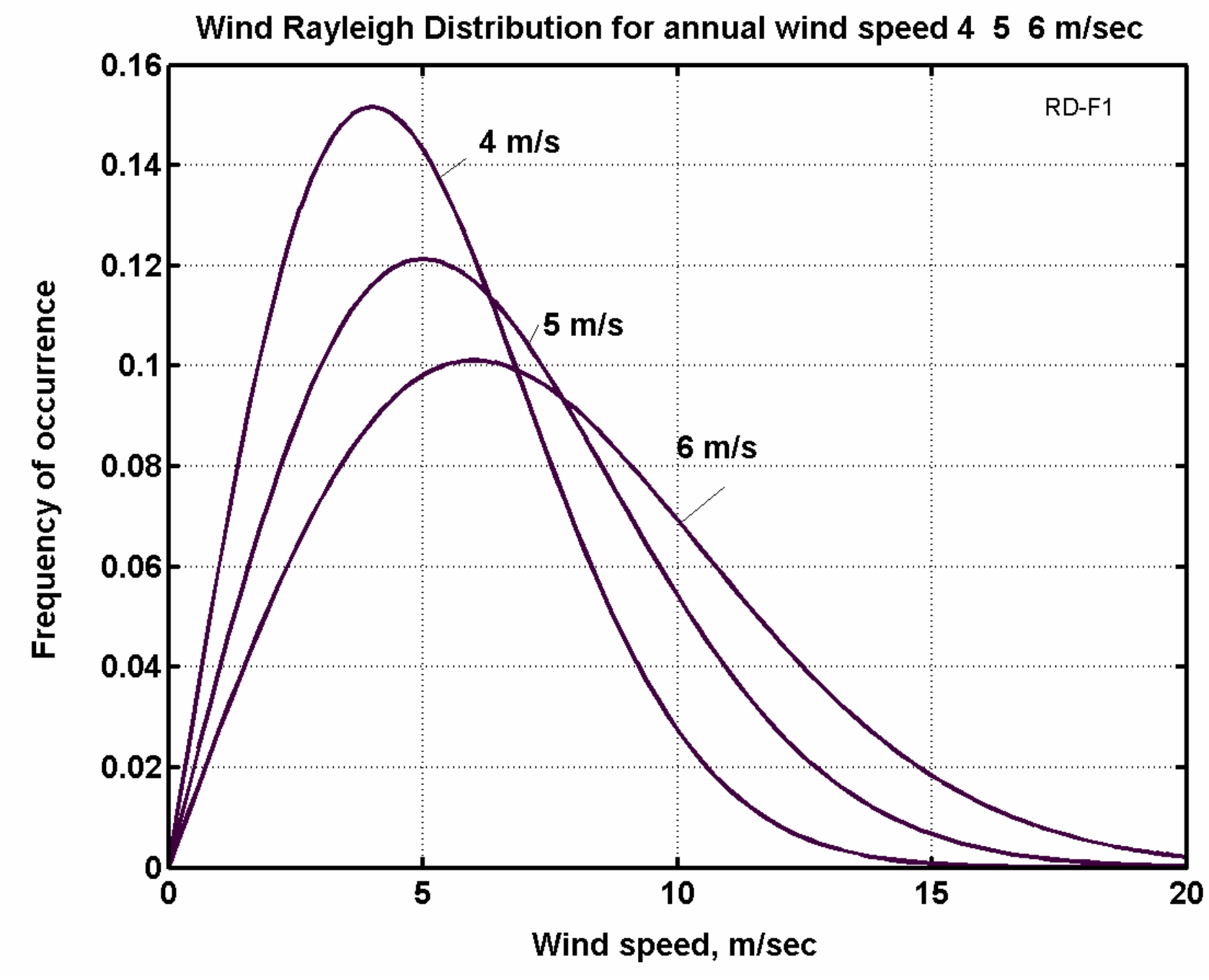

**Fig. 2-3**. Wind speed distribution.

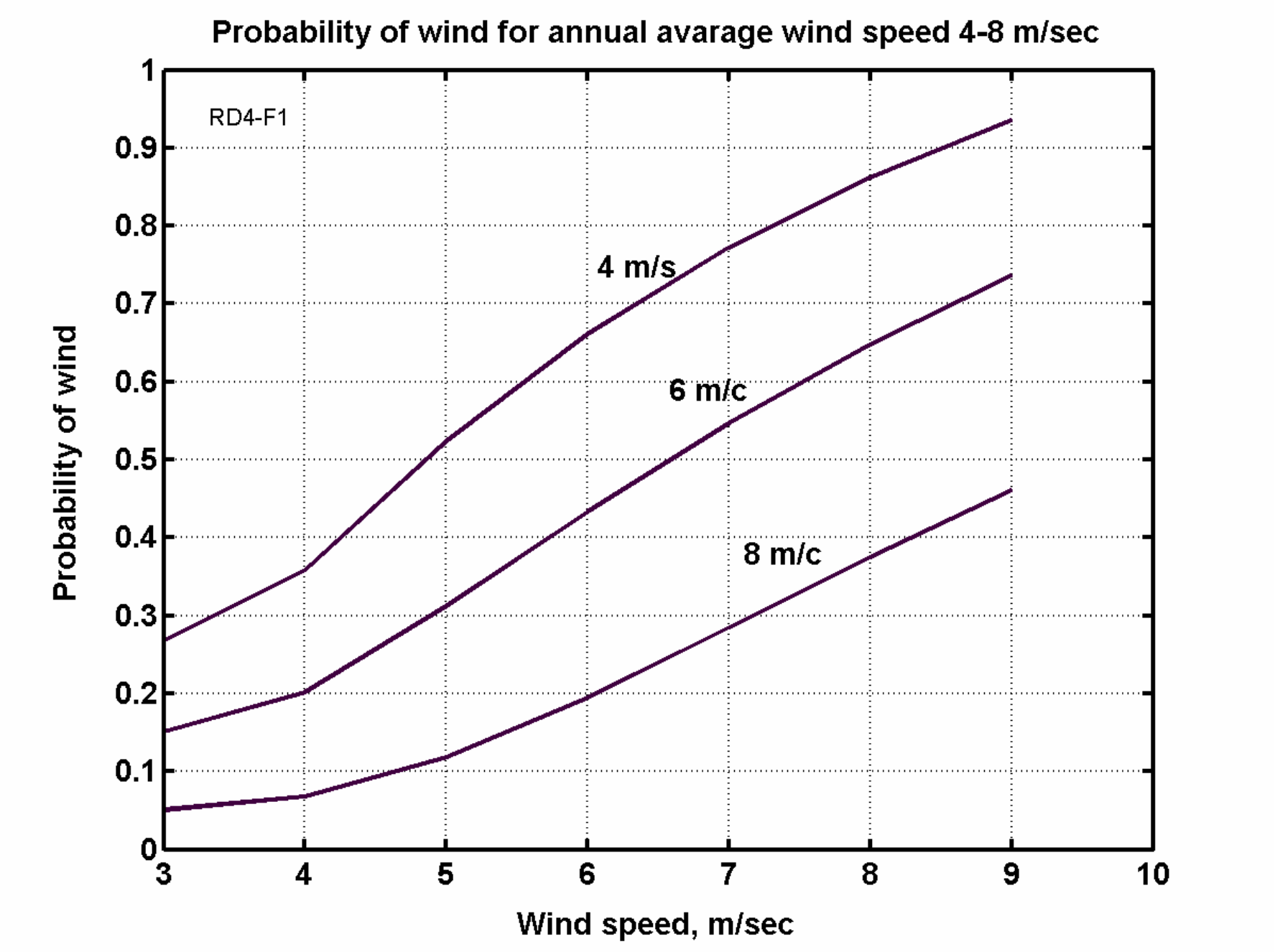

**Fig. 2-3a.**

Let us compute two examples:

1) **MAO**. Assume, the observer has minimum admissible wind speed 3 m/s, maximum admissible speed 25 m/s, altitude 100 m, the average annual speed in given region is 6 m/s. From Table 2-2 and fig. 2-2, 2-3, 2-3a, Eq. (2-1) , we can get the wind speed is 8.4 at $H$=100m, the probability that the wind speed will be less the 2 m/s is 8%, less 3 m/sec is 15%, the probability that the wind speed will be more 25 m/s is closed to 0.

2) **MAV**. Assume, the average annual speed in given region is 6 m/s, the maximum admissible wind speed is 7 m/s. The probability that a wind speed will be less then 7 m/s is 55%, less then 8 m/s is 65%.

It means that the impossible time for MAV flights is 2-3 times more that for kite.

## 2.2. Lift Force

Lift force of kite may be computed by equation

$$L = C_L \frac{\rho V^2}{2} S, \qquad (2\text{-}2)$$

where $L$ -lift force [N]; $C_L$ - lift coefficient, $C_L$ =0 - 2; $\rho$ - air density, if altitude $H$ closed to 0 $\rho$ = 1.225 kg/m$^3$; $V$ - wind speed [m/sec]; $S$ - wing area [m$^2$].

Result of computation for $S$ = 1 m$^2$ and $H$=0 are presented in figs. 2-4, 2-5 (not included); for $S$ = 0.01 m$^2$ and $H$=0 is presented in fig. 2-6 (not included), for $S$ = 1 m$^2$ and $H$=0-8 km are presented in figs. 2-7 (not included).

When designer know load, admissible wind speed, altitude, and fiber stress, he can estimate the necessary wing area.

## 2.3. Cable (Fiber) Mass.

Cable mass can be computed by equation:

$$M = l\frac{\gamma}{\sigma}F, \tag{2-3}$$

where $M$ - mass of cable (fiber) [kg]; $l$ - length of cable [m]; $\gamma$ - cable density [kg/m$^3$]; $\sigma$ - cable stress [kg/m$^2$]; F - tensile force [N].

Result of computation is presented in fig.2-8. As you see the mass of cable is small.

### 2.4. Diameter of the Cable

Diameter of cable (fiber), $d$, may be computed by equation

$$d = 2\sqrt{\frac{F}{\pi\sigma}}\ . \tag{2-4}$$

Results of computation are presented in figs. 2-9, 2-10 (not included).

### 2.5. Drag of the Cable

Drag of main cable (fiber) can be calculated by equation

$$D_{f,1} = C_{D,f1}\frac{\rho V^2}{2}\frac{\pi}{4}ld\ , \tag{2-5}$$

where $D_{f,1}$ -drag of main cable in [N]; $C_{D,f1}$ - drag coefficient; $l$ - length of main cable [m].

Results of computation are presented in fig. 2-11 (not included).

### 2.6. Kite Cable Angle.

The kite cable angle to horizon without cable drag, $\varphi_1$, and with cable drag, $\varphi_2$, may be calculate by equations

$$\tan\varphi_1 = \frac{C_L}{C_D} = \frac{C_L}{C_{Do} + C_L^2/\pi\lambda};\quad \tan\varphi_2 = \frac{C_L}{C_D + 0.5\frac{S_f}{S}C_{D,f1}}, \tag{2-6),(2-7}$$

where $C_{Do}$ - kite drag when $C_L$ =0; $\lambda$- wing aspect ratio, $S_f$ - drag fiber area, $S_f = Hd$ [m$^2$]; S - wing area [m$^2$]; $C_{D,f1}$ - cable drag coefficient.

Result of computation is presented in fig.2-12.

### 2.7. Deviation of Video Cable from Vertical

Video cable angle from vertical may be computed by equation

$$\tan\varphi_3 = \frac{D_{TV} + 0.5D_{f,2}}{g(G_{TV} + 0.5G_{f,2})}\ , \tag{2-8}$$

where

$$D_{TV} = C_{D,TV}\frac{\rho V^2}{2}S_{TV}\ ,\quad D_{f,2} = C_{D,f2}\frac{\rho V^2}{2}(0.75Hd),\quad G_{f,2} = \frac{\pi}{4}\gamma L d_1^2\ , \tag{2-9}$$

$D_{TV}$ - drag of video apparatus [N], $C_{D,TV}$ - drag coefficient of the video apparatus, $S_{TV}$ - reference video apparatus area [sq.m], $D_{f,2}$ - drag of suspended TV fiber [N], $C_{D,f2}$ - drag coefficient TV fiber [N], $H$ - kite altitude [m], $G_f$ - TV cable weight [kg], $\gamma$ - TV cable density [kg/m$^3$], $L$ - TV cable length [m], $d_1$ - diameter of TV cable [m].

Result of computation is presented in fig.2-13.

### 2.8. Full Kite Cable Angle

Full kite cable angle φ may be calculated by equation

$$\tan\varphi = \frac{C_L - g(G + 0.5G_{f,1} + G_{f,2} + G_{TV})/qS}{C_D + 0.5C_{D,f1}\frac{S_{f,1}}{S} + C_{D,f2}\frac{Ld}{S}}\ , \tag{2-9}$$

where $G$ - weight of kite[kg]; $G_{f,1}$ - weight of main cable (fiber)[kg]; $G_{f,2}$ - weight of TV cable [kg]; $G_{TV}$ - weight of video (TV) camera (apparatus); $q = \rho V^2/2$ - dynamic pressure [n/sq.m]; $S$ - wing area [sq.m]; $C_{D,f1}$ - drag coefficient of main cable; $S_{f,1} = Hd_1$ - reference video cable area.

Results of computation are presented in fig. 2-14 (not included). Here are $G$ = 0.2 kg, $G_{TV}$= 0.03 kg.

## 2.9. Cable Thrust

Thrust of main cable may be computed by equation

$$T = \frac{qS}{g}\sqrt{T_t^2 + T_2^2}\,, \tag{2-10}$$

where

$$T_1 = C_D + 0.5C_{D,f1}\frac{S_{f,1}}{S} + C_{D,f2}\frac{Ld}{S}\,,$$
$$T_2 = C_L - g(G + 0.5G_{f,1} + G_{f,2} + G_{TV})/qS\,. \tag{2-11}$$

Result of computation is in fig.2-15. Here are $G$ = 0.2 kg, $G_{TV}$= 0.03 kg.

## 2.10. Viewing Distance (distance of video signal).

The distance L which can be viewed of the Earth from a high altitude (antenna) is given by

$$L = \sqrt{2R_e H + H^2}\,, \tag{2-12}$$

where Re=6378 km is the Earth radius, H is an antenna altitude. The results of computation are presented in fig.2-16. As us see the MAO and MAO can transfer video signal in distance of hundreds times more then current MAV, which has range only 0.3 – 1 km (see Attn. # ).

## 2.11. Mass and Admissible Current of Wire

The admissible current in wire depends from relation a gross-section area to a cooled wire surface. That why it linear depends from a diameter of wire. For aluminum and cupper wire these ratio are following respectively:

$$J_1 = 8d, \quad J_2 = 10d\,, \tag{2-13}$$

where $J_1$, $J_2$ are admissible current (ampere), d is wire diameter [mm]. Result of computation is in fig.17.

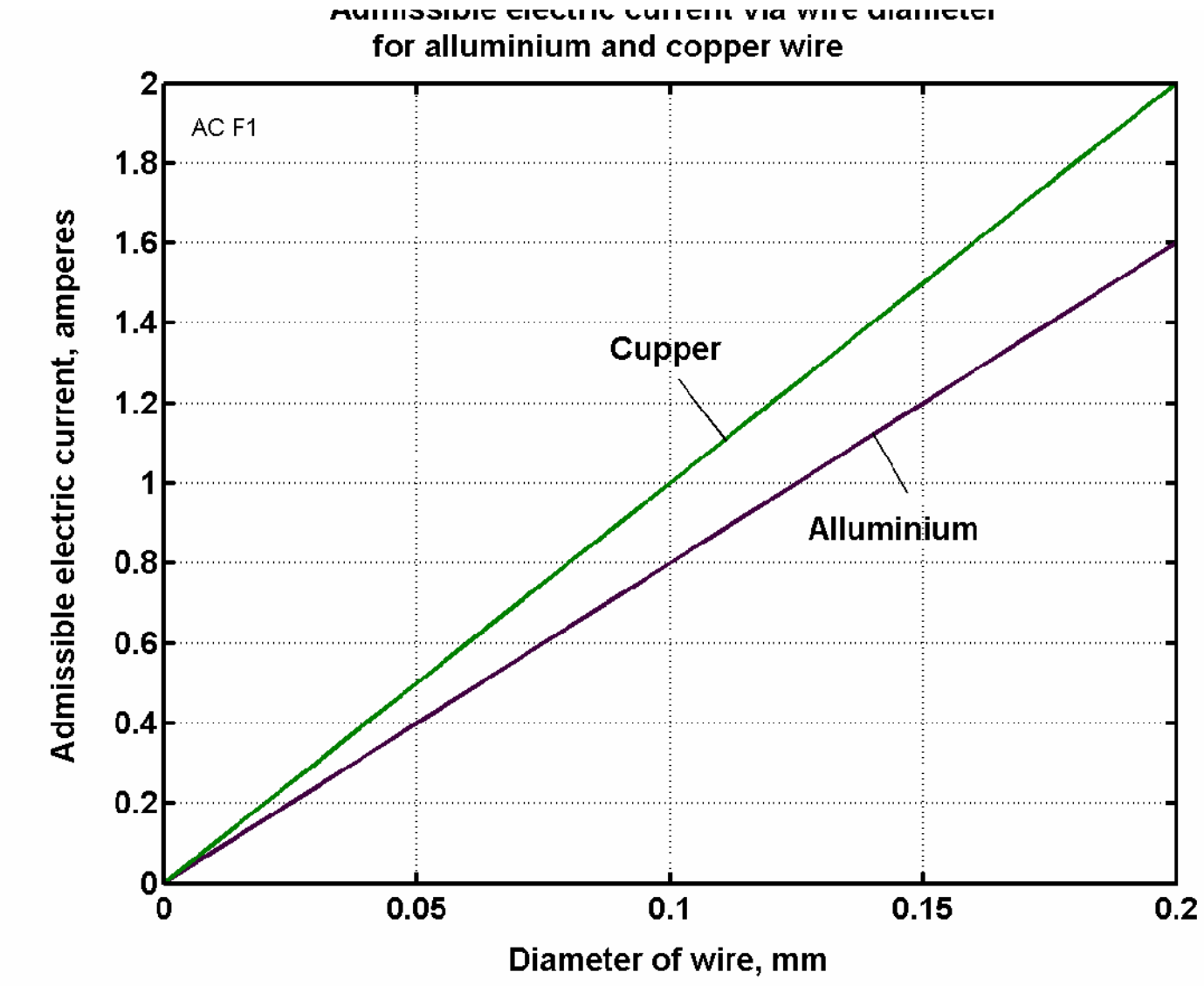

**Fig. 2-17.** Safety electric current via wire diameter for alluminium and cooper wire.

The weight, $W$, [g] of wire is respectively

$$W_1 = \frac{\pi}{4} d^2 \gamma_1 L\,, \quad W_2 = \frac{\pi}{4} d^2 \gamma_2 L\,, \tag{2-14}$$

where d is wire diameter [sm], $\gamma$ - density [g/cm3], L is a wire length [cm]. For aluminum $\gamma$=2.7 g/cm3, for copper $\gamma$=8.93 g/cm3. The result of computation for l=100m is presented in fig.2-18.

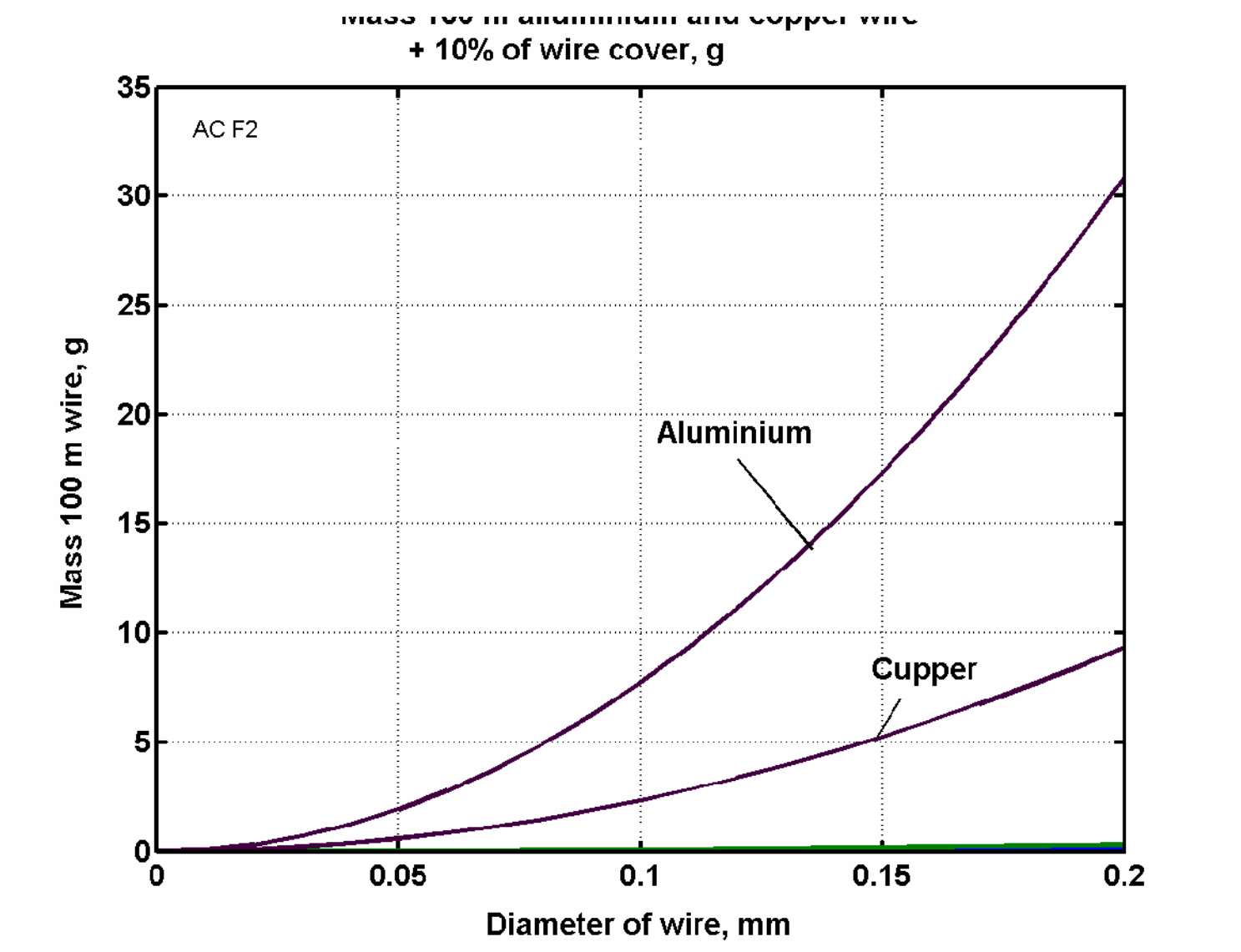

**Fig. 2-18**. Mass 0f 100 m aluminum and cooper wire + 10% of wire cover (isolator), g.

### 2.12. Lift force of Air Balloon.

Let us to compute lift force, mass of cover, and useful lift force of air balloon filled by helium and hydrogen. Assume that a balloon length equals three balloon diameter.

$$V = \frac{3\pi}{4} d^3, \quad L = (\gamma_a - \gamma_g)V, \quad S = 5\pi d^2, \quad M = S\delta\gamma, \quad L_u = L - M, \tag{2-15}$$

where $V$ is balloon volume [m3], $d$ is the balloon diameter [m], $L$ is the lift force [kg], $L_u$ is useful lift force (without balloon cover); $\gamma_a$, $\gamma_g$, $\gamma$ are density of air, gas, cover respectively: $\gamma_a$ =1.225 kg/m$^3$ for air, $\gamma_g$= 0.1785 kg/m$^3$ for helium, $\gamma_g$= 0.09888 kg/m$^3$ for hydrogen, and $\gamma$ =1800 kg/m$^3$ for balloon cover, $S$ is area of cover [m2]; $M$ is mass of cover [kg].
Result of computation is presented in figs. 2-19 – 2-22.

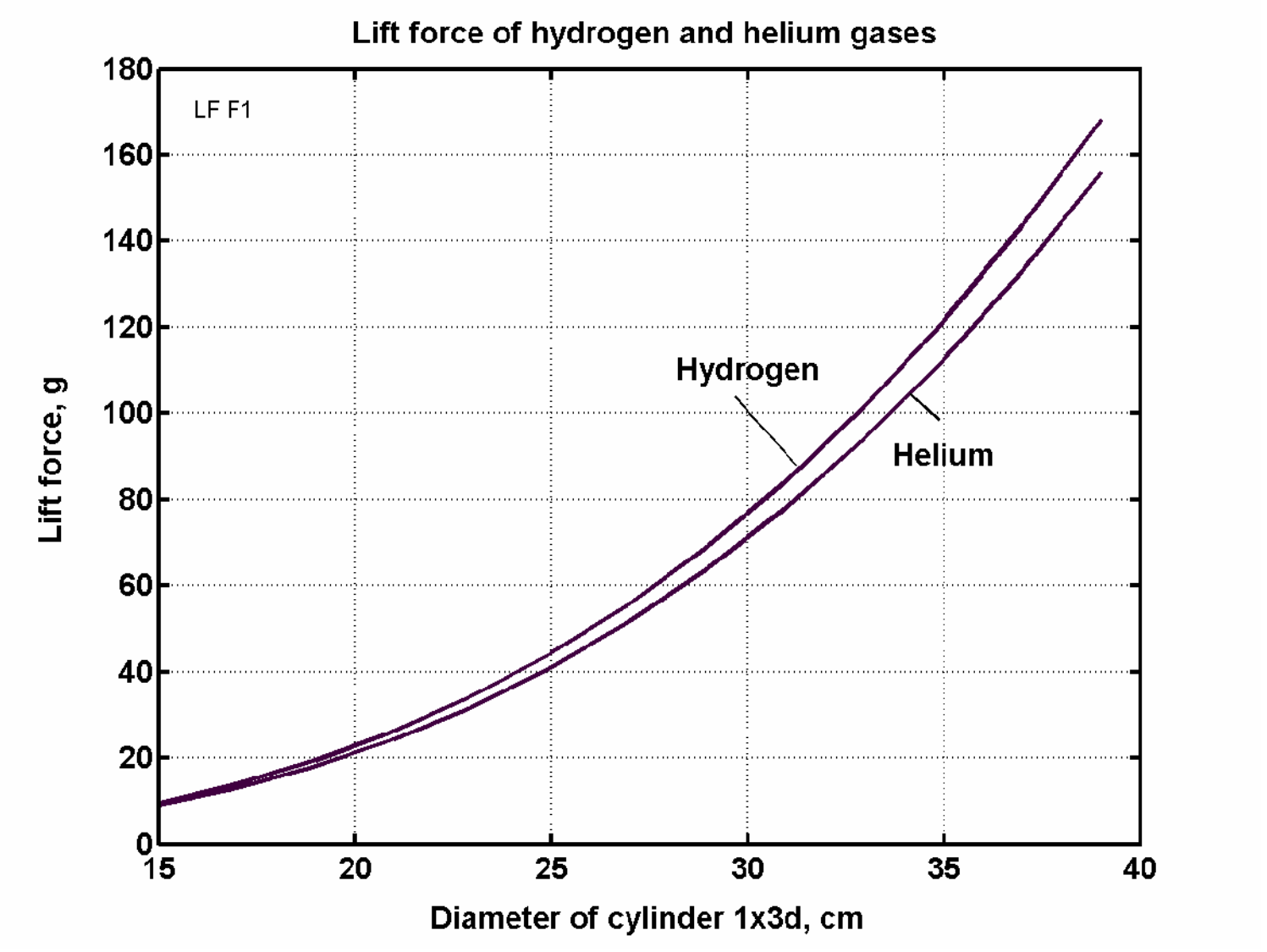

**Fig. 2-19**. Lift force of balloon filled the hydrogen and helium gases.

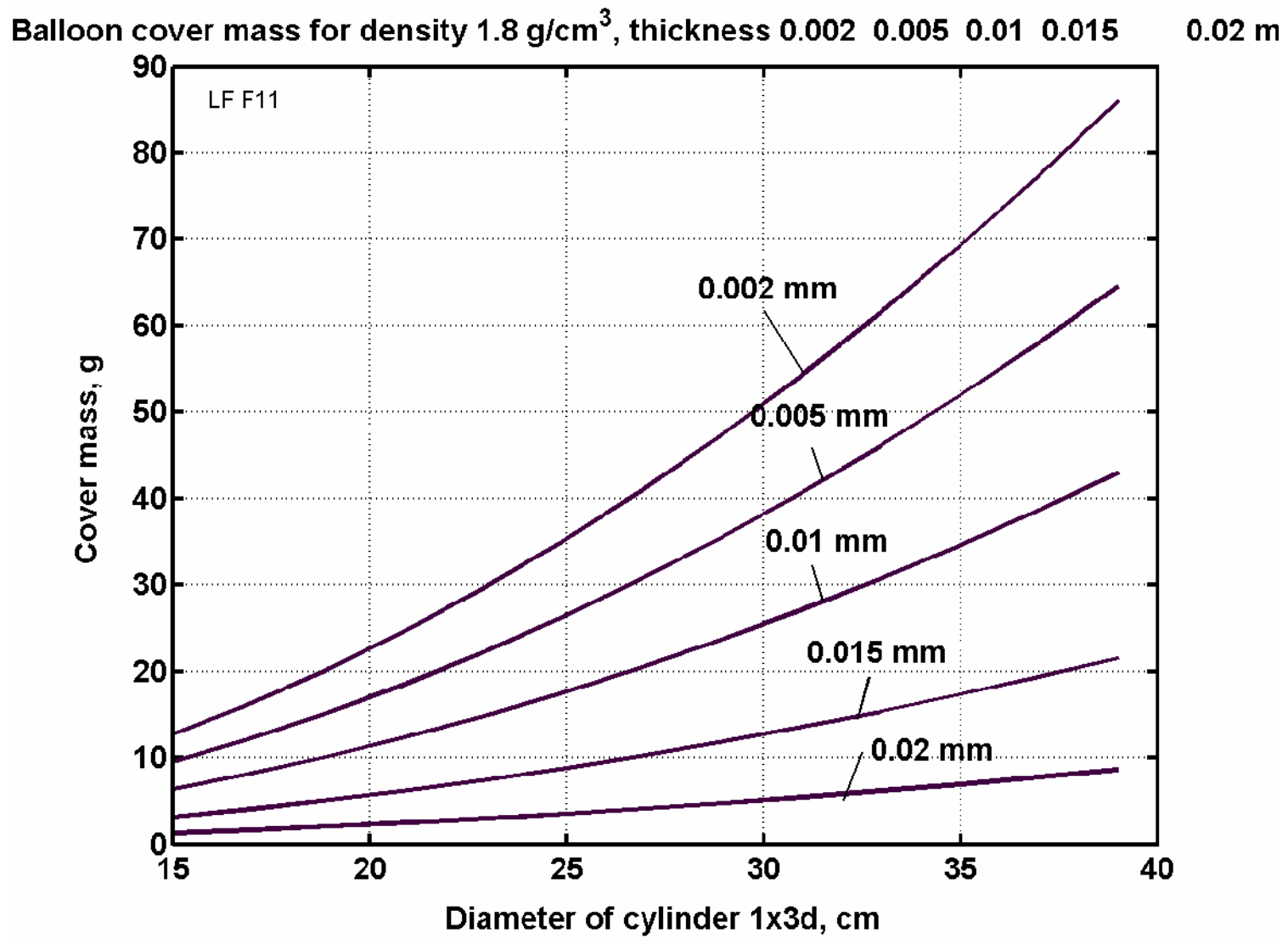

**Fig. 2- 20.** Balloon cover mass for cover density 1.8 g/cm$^3$, cover thickness 0.002, 0.005, 0.01, 0.015, 0.02 mm..

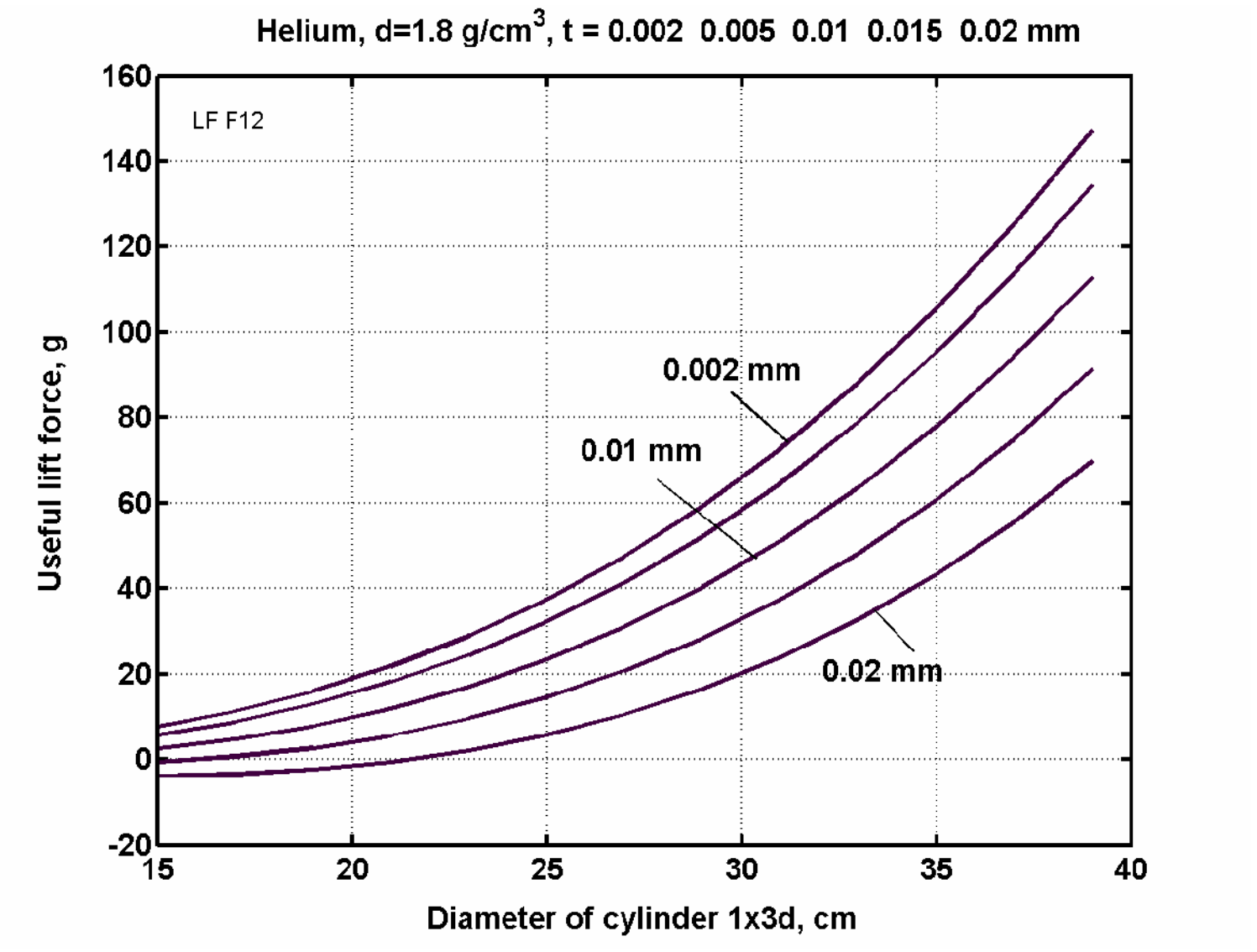

**Fig. 2-21**. Usefull lift force of helium balloon 1x3d for cover density 1.8 $cm^3$, cover thickness 0.002 0.005 0.01 0.015 0.02 mm.

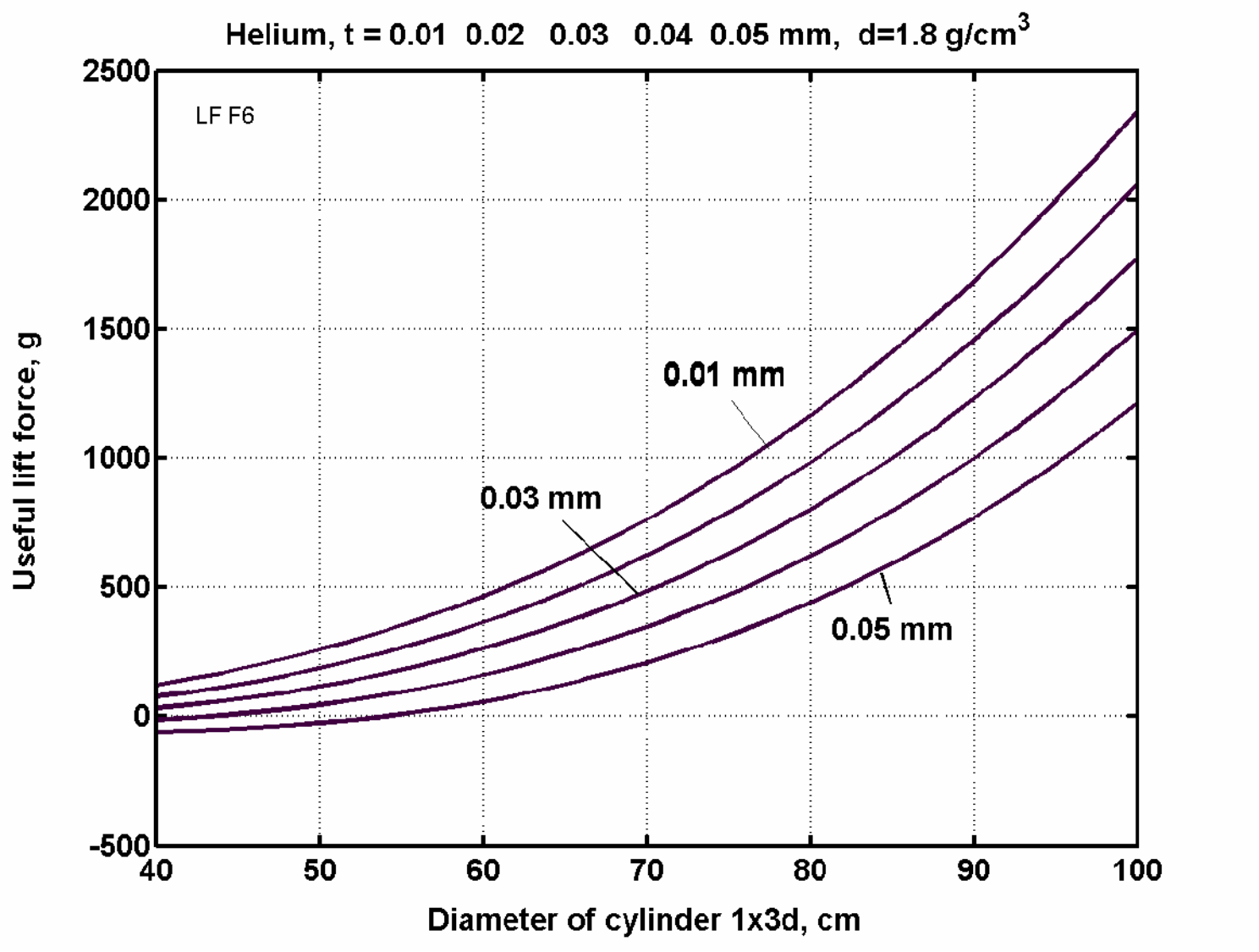

**Fig. 2-22.** Useful lift force of helium via cylinder diameter for cover thickness 0.01 0.02 0.03 0.04 0.05 mm, cover density 1.8 $g/cm^3$.

# 3. Requirement Video Camera and Control

The capabilities of video camera are very important component for MAV, MAO and MAO. The video camera must be small, light as soon as possible. One must recognizes a target from maximum (as soon as possible) distance. It means, the TV camera must have a maximum (as soon as possible) pixels. It will good, if Camera has a zoom or small FOV degree.

There are Johnson Criterion to estimate $a$=66+/-12 pixels across the minimum dimension of a target for 99.9% to 99.999% probability of ID.

Some target sizes (*l* x *w* x *h* in meter):

- M113 Typical armored personnel carrier: 4.7 x 2.5 x 1.8 m.
- M1 Main Battle Tank: 7.9 x 3.7 x 2.4 m.
- Scud surrogate George: 13.3 x 3.0 x 2.5 m.
- Man: 0.5 x 0.5 x 1.7 m.
- Man face: 0.3 x 0.3 x0.3.

### 3.1. Required Number of Pixels

The target size *s* [m], view angle $\alpha$ [degree], distance to target *D* [m], pixel number *P* (for one side, full is *P* x *P*), number of identification pixels *a,* are described by equation

$$a = \frac{57.3sP}{\alpha D} \quad . \qquad (3\text{-}1)$$

Result of computation of pixels *p* via distance for *s* =3 m, *P* =1000 and 2000 pixels, $\alpha = 1^o$-$50^o$ degrees are presented in firs.3-1, 3-2 (not included).

### 3.2. Recognition Distance and Target Size

Recognition distance via the target size may be computed by equation below derived from eq. (3-1)

$$D = \frac{57.3sP}{\alpha a} \quad . \qquad (3\text{-}2)$$

Results of computation of recognition distance *D* via target size *s* for 256-2000 FAO pixels *P* and α= 1, 2 5, 20 degrees, are presented in figs. 3-3, - 3-6 (not included).

Results of computation of recognition distance *D* via target size *s* for 1000 FAO pixels *P, a*=66*,* and α= 1, 2 5, 20, 50 degrees, are presented in fig. 3-7 (not included).

As you see, the current TV cameras request a low flying of MAV. It is bad for MAV because an enemy sees MAV and the enemy can put out of sight or annihilate MAV. The MAO and MAO also need in low location of video apparatus. However it is well, because video camera is separated from MAO, the camera and fibers are small and they may be difficult for enemy recognizing. That way we need to separate video camera from MAO or MAO.

### 3.3. Communication problems

Communication problems for MAV relate primarily the bandwidth required to the small video size, hence small antenna size, and to the limited power available to support the bandwidth required (2-4 megabits per second) for image transmission. Control functions demand much lower bandwidth capabilities, in the 10's of kilobits range, at most. Image compression help reduce the bandwidth requirement, but this increases on-board processing and hence requirements. The limited power budget means the omni-directional signal will be quite weak. So directional ground antennas may be required to track the vehicle, using line-of-sight transmissions. But limitations to line-of-sight would be severely restrictive for urban operations, so architectures.

Most these problems absent for MAO and MAO or do not have these strong limitations. The MAO do not have strong limitation in size and weight because the main kite is located at high altitude and video apparatus do hot have wings. The lower apparatus can has a video camera size. No big limits in electric power, because no strong limit in weight. The fiber can has thin wires and the electricity can transfer from soldier to video camera. The kite is located at high altitude and the fiber-wires can be used as antenna. The signal distance (range) is increased in a lot of times. The MAO, MAO can make the surveillance permanently in jungle and small yard between buildings, in a given window.

If MAV has speed 20 m/sec and operator needs 3 sec to see picture, to understand image, and to control of MAV, it means the MAV cannot to fly in area having size less 60 m. In reality this size more. For example, if a MAV turn time is 3 sec, we must have additional 60 meters for turning (total 60 + 60 =120 m). It means the MAV cannot operate in city having a compact development.

# 4. Examples

### 4.1. Example #1. (Four anchor MAO launched from aircraft). Main parameters.

**Kite:**

Wing area 0.08-0.12 $m^2$

Minimal wind speed 3 m/sec. Minimal lift force for minimal wind speed 90-130 g.

Mass: Kite 20 – 30 g; TV – station 20 g; fiber cable (include wire) of diameter $d$ =0.1-0.15 mm, of the total length 1000 m has mass 20-32 g. Total mass: 70 – 90 g (without anchors and battery located at anchor).

Admissible electric current is up 0.4 amperes for wire diameter 0.05 mm.

Operative kite altitude: 100- 200 m. (may be up 500 m).

**Video station:**

Minimal operative altitude of video station 5-7 m.

Permanently observe area (100-400) x (100-400) m.

Data of video camera and microphone:

Mass 8-20 g. Size 2-8 cm.

Maximum recognize distance, $D$, of targets size 2 m:

1. For FOV angle $\alpha=20^0$, pixels $P$=256, $D$ is 22 m;
   For pixels 1000, $D$ is 87 m.
2. For FOV angle $\alpha=2^0$, pixels $P$=256, $D$ is 220 m;
   For pixels $P$=1000, $D$ is 850 m.

**Probability** of wind more 3 m/sec in area with average annual speed 6 m/s at altitude 10 m is 0.85. In our case this probably is more (about 0.9) because our kite located at altitude 100-200 m. Permanently operation time is some weeks or months.

### 4.2. Example #2. (Balloon MAO launched from aircraft). Main parameters.

**Wing balloon:**

Diameter of cylindrical 1x3 balloon with useful lift force 115 g and cover thickness 0.01 mm is $D$=39 cm . Wing area $S$=0.06-0.1 sq.m

Minimal wind speed: None. Maximum useful (without balloon cover and zero wind speed) lift force is 115 g.

**Mass**: Balloon cover of thickness 0.01 mm - 42 g; video – station 20 g; fiber cable (include wire $d$=0.05 mm) has diameter $d$ =0.1-0.15 mm, maximum tensile force is 6-12 kg, total length 1000 m, mass 20-32 g. Total mass: 70 – 90 g (without anchors and battery located at the anchor).

Diameter electric wire is 0.05 mm. Admissible electric current is up 0.4 amperes.

Operative balloon altitude: 100- 200 m. (may be up 500m)

**Video station:**

Minimal operative altitude of video station 5-7 m.

Permanently observe area (100-400) x (100-400) m.

Data of video camera and microphone:

Mass 8-20 g. Size 2-8 cm.

Maximum recognize distance, $D$, of targets size 2 m:

3. For FOV angle $\alpha=20^0$ , pixels $P$=256, $D$ is 22 m;
   For pixels 1000, $D$ is 87 m.
4. For FOV angle $\alpha=2^0$, pixels $P$=256, $D$ is 220 m;
   For pixels $P$=1000, $D$ is 850 m.

**Probability** of wind more then admissible maximum speed 15 m/sec in area with average annual speed 6 m/s at altitude 10 m is very small (≈0.01). In our case this probably is more (about 0.02) because our balloon located at altitude 100-200 m.

Permanently operation time is some weeks or months.

### 4.3. Example #3 (Kite MAO for soldier). Main parameters.

Soldier **kite** apparatus:

Wing area 0.05- 0.08 $m^2$.

Weight 50-150 g.

Minimum speed 3 m/sec. Probability is 0.85 for area with the average annual wind speed 6 m/sec.

Operative kite Altitude 100- 150 m.

Permanently observe area (100-500) x (30-200) m.

**Video camera** and microphone apparatus:

Weight 9-25 g.

Size 2-8 cm.

Operative altitude 7-50 m.

Maximum recognize distance, $D$, of targets size 2 m:

5. For FOV angle $\alpha=20^0$ , pixels $P$=256, $D$ is 22 m;
   For pixels 1000, $D$ is 87 m.
6. For FOV angle $\alpha=2^0$, pixels $P$=256, $D$ is 220 m;
   For pixels $P$=1000, $D$ is 850 m.

Operative time limited by battery located at anchor (may be same hours).

### 4.4. Example #4. Munitions Air Observer (MuAO). Main parameters.

Main wind (**kite**) MuAO apparatus:

Wing area 2-8 sq.m. Minimum wind speed 3-4 m/sec.

Weight 20-100 kg

Operative Altitude 500- 2000 m.

Observe area up 2 x 2 km.

Number of anchors: 4 – 6.

Arming: number of control anti-tank projectile is 5-20 (2-3 kg each),

Number of control small anti-man grenade is 10-50 (0.1-0.3 kg each).

**Video cameras**:

Number 4 – 12.

Weight (each) 25-100 g.

Size 5-12 cm.

Operative altitude 20-100 m.

Maximum recognize distance, $D$, of targets size 2 m:

7. For FOV angle $\alpha=20^0$ , pixels $P$ =256, $D$ is 22 m;
   For pixels 1000, $D$ is 87 m.
8. For FOV angle $\alpha=2^0$, pixels $P$ =256, $D$ is 220 m;
   For pixels $P$ =1000, $D$ is 850 m.

Operative time: some weeks.

Probability of wind is about 0.9 at this altitude in area with average wind speed 6 m/sec

# Attachment #1:

## 1. Plan of Future MAO Research and Development.

**Researches:**

1. Finding of information weight, volume, and other technical parameters of current devices (small video camera, video transmitter, TV receiver, battery, radio control, fiber), which can be suitable for the offered MAO, MAO.
2. Studying the current video equipment of very small (soldier) recognizer unmanned aircraft (MAV) and possibility their application for MAO, MAO, missiles, bombs, and gun shells.
3. Estimation of cost (in case of a widely producing or big order) the device necessary for MAO, MAO, civil industry (police, emergency agency), especially the cost of widely produced very small video cameras. Possibility their size decreasing and improving of the technical parameters in future.
4. Computation of the main parameters.
5. Schematic design of the MAO and solution of the main problems, which can appear in the offered MAO. Design Airframes, actuation, control laws.
6. Patenting the offered method and device (MAO).
7. Publication (?).
8. Announced of prizes for better MAO in aviation modelers.
9. Testing best MAO.

**Development:**

1. Detailed design and manufacturing 5-10 the best MAO for wide testing.
2. Testing MAO as observation device.
3. Testing MAO by tube (or gun) launcher.
4. Real testing in army.

**Application:**

1. The order for industry.
2. Widely application in military operation.
3. Widely application in civil life: for observe car traffic by police, area after disaster, in rescue operation, and so on.

# Attachment #2:

## The short history of MAV R&D.

The recent history of MAVs start with a 1992 workshop on future technologies for military operations held by DARPA at Rand. Then-senior scientist Augenstein led a panel on micro vehicles, including aircraft systems ranging in size from a hummingbird down to less than 1 cm. Rand published a widely circulated report on the work in 1994. The Lincoln Laboratory was initially skeptical, but its own research also concluded that MAVs were becoming feasible.

DARPA held a MAV feasibility workshop in November 1995, a briefing to industry in March 1996, and a user and developer workshop in October 1996. These were mainly paper exercises with little real hardware. The Lincoln Laboratory conducts studies and the Naval Research laboratory acts as a technical agent for DARPA.

In Fiscal 1997, DARPA started a $35-million, four-year effort to develop and demonstrate affordable MAVs. The agency wants aircraft with a maximum dimension of 6 in. (152 mm) to fly ranges up to 10 km and speeds up 40-50 mph. (70-80 km/hour) for missions that are 20 min to 2 hour long. MAVs are to be deployed by hand, by munitions launch or from larger aircraft.

Missions include reconnaissance, targeting, placing sensors, communications relay and sensing dangerous substances. They are viewed as one-use, one-way missions.

DARPA's Tactical Technology Office awarded nine Phase 1 small business innovative research (SBIR) contracts worth $100,000 each to either pursue system development or a specific technology.

Four of the contracts ($750,000 each) progressed into Phase 2 in Fiscal 1998.

After DARPA this problem are trying to solve AF and Army. Unfortunately, after 10 years R&D no MAV that can fly to back size of building and show what is located behind of the building.

## Attachment #3

**Main parameters of some current video, radio Control and other devices.**

Video Components:
Camera –2.4 ghz camera and transmitter Model-CMDX-22): www.rf-links.com
Receiver – 2.4 ghz receiver (model – VRX-241t): www.rf-links.com
Receiver – Extreme 5/M 5: www.fmadirect.com
Antenna – High Gain 2.4 ghz panel antenna (Model – PN-24S): www.rf-links.com
Transmitter/Controller – Futaba 9CAP digital controller: www.towerhobbies.com

The manufacturer claims a range of the video camera transmitter and receiver a 3000 ft. line of sight range. The radio equipment has a range to 1500 m line of sight. Resolution about 330 lines. Pinhole lens.

Video camera model: CMX-916. Price: $329.

DATA: Battery operated 9 V-12 V; current consumption 68 mA/9 V, 82 mA/12 V; RF Power 80 mW; Size 0.7"x 0.7"x 0.8"; Weight 8 grams; Range from 300 ft up to 3000 ft LOS; Color picture, no audio; frequency 916 MHz.

Receivers: Models: PTU-402 price $159, VRX-24 L price $299, weight 300 g

1. **Minicam** (fig.A3-1)(http://www.helihobby.com/html/micro_video_camera.html)
   The "minicam" all-in-one color video camera with built in transmitter available. It is utilizing 2.4 GHz technology. The minicam weight 1/3 oz (9 grams) and comes complete with a color camera, transmitter, and receiver. Price is $260. (info@helihobby.com).
2. **Eyecam** on-board wireless camera (http://www.reallycooltoys.com/toys/i4info.html)
   This all-in-one color video camera with built in transmitter available. It is utilizing 2.4 GHz technology. Weight 9 grams. Camera and transmitter size: 15mm x 22mm x 32mm. Camera Lux: <3f1.2. Camera Auto Electronic Exposure of 1/60 to 1/15000 sec. Camera pixel resolution is 365K (PAL) or 250K (NTSC). Wireless Transmission Range: 300 M (1000'). Complete has color camera, transmitter, and receiver. Price is $395.
3. **Wireless Micro Video camera**. http://www.pimall.com/nais.nl/n.thirdy.html. Wireless transmission is 434 MHz, 900 MHz, 2.4 GHz.. Range 1500-3500 foot.

## Attachment #4:

Industry electric Model RC Helicopters: www.hobby-lobby.com/elecheli.htm
Industry Electric Airplanes: www.hobby-lobby.com/slowflyers.htm .
Industry Electric Flight Accessories: www.hobby-lobby.com/eflight.htm

**Main parameters of the best current MAV**

1.**AeroVironment** (Aviation Week, June 8, 1998, p.47.) Fig. A3-1.The company wants to reach:

a) Line of Sight Operation 1 km. Radius.
b) 10 min. Duration

The current model has:

c) Black & White Video Payload.
d) Size: disk 6", 152 mm

Table A-1 (Company are wanted)

| Aircraft Subsystem | Weight (grams) | Peak Power (mW) | Average Power (mW) |
|---|---|---|---|
| Lithium battery | 26 | 0 | 0 |
| Propulsion Motor | 7 | 4000 | 2000 |
| Gearbox | 1 | 0 | 0 |
| Propeller | 2 | 0 | 0 |
| Airframe | 4 | 0 | 0 |
| Control actuator | 1 | 200 | 200 |
| Receiver & CPU | 1 | 50 | 50 |
| Downlink Transmitter | 3 | 1200 | 300 |
| B/W Video Camera | 2 | 150 | 50 |
| Interface electronic | 1 | 50 | 50 |
| Roll Rate Gyro | 1 | 60 | 60 |
| Magnetic Compass | 1 | 180 | 180 |
| Total | 50 | 5800 | 2890 |

2. **MicroStar** of Lockheed Martin (Aviation Week, November 9 1998, p.37.Fig.A3-2.
The company want to built MAV:
Long 6 in, weight 3 oz, flight duration 20 min, cover distance up to 3 mi., and cruise at 30 mph. Takeoff weight is 86 grams: 18 g payload, 9 g airframe, 44.5 g power source, 13.5 g tealthy electric engine.

Cost from $5,000-10,000 each. Day/night Camera 512 x 512 pixels, 30 frames per sec. Flight altitude 150-500 ft.

Operations would be limited by winds of 30 mph. Or more, fog, heavy dust and rain. Wind could be particularly onerous in cities where buildings produce micro bursts that might bring down the UAV.

## Attachment #5. Main parameters of industry produced artificial fiber

Cable discussing. Let us to consider the following experimental and industrial fibers, whiskers, and nanotubes:

1. Industrial fibers have $\sigma$ = 500-620 kg/mm$^2$, $\gamma$ =1560-1950 kg/m$^3$, for $K$=$\sigma$/$\gamma$/2.4 =600/1800/2.4= 0.139. Young,s modules of graphite fiber is up 200 Gpa, aluminum is 63-70 Gpa, Cupper is 127 Gpa.
2. Whiskers $C_D$ has $\sigma$ = 8000 kg/mm$^2$, $\gamma$ = 3500 kg/m$^3$ (1989)[2], p.158. We can take for computation $\sigma$ = 8000/2.4=3333 kg/mm$^2$, $\gamma$ = 3500kg/m$^3$, $k$ =$\sigma$/$\gamma$ = $9.5\cdot10^6$, $K$=0.95.
3. Experimental nanotubes CNT (Carbon nanotubes) have tensile strength 200 Giga-Pascals (20000 kg/mm$^2$), Young's modules is over 1 Tera-Pascal, specific density $\gamma$=1800 kg/m$^3$ (1.8 g/cc)(2000 year). For safety factor $n$ =*2.4,* $\sigma$ = 8300 kg/mm$^2$=$8.3\cdot10^{10}$ n/m$^2$, $\gamma$ =1800 kg/m$^3$, $k$ =$\delta$/$\gamma$ =$46\cdot10^6$ ($K$=$10^{-7}$k=4.6). The theory predicts 1 Tera Pascals and Young modules 1-5 Tera Pascals. The nanotubes SWNT's have density 0.8 g/cc, the nanotubes MWNT's have the density 1.8 g/cc.

The reader can find the cable discussing in [1] and cable characteristics in [2]-[5]. In our projects we use only current cheap artificial fibers widely produced by current industry.

*References of artificial fiber, whisker, nanotubes:*

## **Attachment #6. Average Wind Resources of World** (non included)